\documentclass[journal]{IEEEtran}
\usepackage{caption}
\usepackage{color}
\usepackage{ragged2e}
\usepackage{float}
\usepackage{array}
\usepackage{makecell}
\usepackage[linesnumbered]{algorithm2e}
\usepackage{url}
\usepackage[cmex10]{amsmath}
\usepackage{multirow}
\usepackage{setspace}
\usepackage{graphicx}
\usepackage{subfigure}
\usepackage{times}
\usepackage{amsfonts}

\ifCLASSINFOpdf
\else
\fi
\begin{document}
%
\title{Personality-Aware Probabilistic Map for Trajectory Prediction of Pedestrians}
%
%
%

\author{
        Chaochao~Li,
        Pei~Lv,
        Mingliang~Xu,
        Xinyu Wang,
        Dinesh Manocha,
        Bing~Zhou,
        and Meng~Wang

{
 \IEEEcompsocitemizethanks{
 \vfill{Corresponding author: Mingliang Xu}
 \IEEEcompsocthanksitem \vfill{Chaochao Li, Pei Lv, Mingliang Xu, Xinyu Wang, and Bing Zhou are with Center for Interdisciplinary Information Science Research, Zhengzhou University, Zhengzhou 450000, China (e-mail: zzulcc@gs.zzu.edu.cn; \{ielvpei, iexumingliang, iebzhou\} @zzu.edu.cn; wangxinyu@gs.zzu.edu.cn).}
 \IEEEcompsocthanksitem \vfill{Dinesh Manocha is with Department of Computer Science and Electrical $\&$ Computer Engineering, University of Maryland, College Park, MD, USA (e-mail: dm@cs.umd.edu).}
 \IEEEcompsocthanksitem \vfill{Meng Wang is with the School of Computer and Information, Hefei University of Technology, Hefei 230009, China (e-mail:
 eric.mengwang@gmail.com).}
}
}

}

\maketitle

\begin{abstract}
We present a novel trajectory prediction algorithm for pedestrians based on a personality-aware probabilistic feature map. This map is computed using a spatial query structure and each value  represents the probability of the predicted pedestrian passing through various positions in the crowd space. We update this map dynamically based on the agents in the environment and prior trajectory of a pedestrian.  Furthermore, we estimate the personality characteristics of each pedestrian and use them to improve the prediction by estimating the shortest path in this map. Our approach is general and works well on crowd videos with low and high pedestrian density.
We evaluate our model on standard human-trajectory datasets. In practice, our prediction algorithm improves the accuracy by $5-9\%$ over prior  algorithms.
\end{abstract}


%
\IEEEpeerreviewmaketitle

\section{Introduction}
%
%
%
%
\IEEEPARstart{T}{he} problem of modeling or estimating the trajectories of pedestrians in a crowded scene has been studied in computer vision, robotics, augmented reality and other applications. In many applications such as human-robot interaction, autonomous driving or surveillance, it is important to accurately predict the trajectories for collision-free navigation or abnormal behavior detection \cite{2019061301,2019061302}. Typically, the motion of each pedestrian is governed by various factors, including its immediate goal, the environment or the context~\cite{1}, intrinsic personality characteristics~\cite{3,4,13} as well as the movement behaviors of other pedestrians in a crowd~\cite{5,7,8}. One  major challenge is to integrate these complex influencing factors in an intuitive and efficient manner and develop a general scheme that works on most of crowd videos.

There is extensive work on trajectory prediction and different techniques based on tracking filters, motion modeling, or machine learning have been proposed~\cite{32,36,38}. However, the accuracy of these methods varies based on the environment, pedestrian density, and the context (e.g. cultural factors or public settings).
Furthermore, the differences in individual personalities are either ignored or partially accounted by existing trajectory prediction methods. For example, current state-of-the-art machine learning methods based on LSTM neural network do not account for individual personality~\cite{11}. On the other hand, prior research in psychology literature suggests that  individual personality plays a vital role in pedestrian movement~\cite{13}.

\noindent {\bf Main Results:} We present a novel trajectory prediction algorithm for pedestrian movement in a crowd based on a personality-aware probabilistic feature map. This map is computed according to the prior trajectory of each pedestrian and the surrounding scene information. We use a simple spatial query structure based on an equal-sized grid and each grid cell stores a weight that represents the possibility of the predicted pedestrian passing through that cell. In addition, we formulate the relationship between individual personality and the pedestrian trajectory. The individual personality features are extracted from  prior trajectories and fused into the map. Overall, all the influencing factors are combined together in our probabilistic feature map. The final prediction result is obtained by estimating the shortest path in this map. The novel components of our work include:

\begin{itemize}
\item A novel personality-aware probabilistic feature map for individual trajectory prediction based on crowd movements, which accounts for different influencing factors including individual personality.
\item A new feature extraction method to quantify the relationship between the personality of an individual and his or her movement trajectory. The personality features of each pedestrian are estimated from prior historical trajectories and integrated into the probabilistic feature map to predict individual trajectories better.
\end{itemize}

Our method has been evaluated on two public datasets: ETH \cite{48} and UCY \cite{49}, which consists of $1536$ pedestrians in multiple crowd videos with various challenging scenarios.
Compared with  state-of-the-art prediction algorithms, 5-9\% improvement is observed in accuracy using our approach.

\section{Related work}
There is considerable work on trajectory prediction in AI, computer vision, and robotics. Some simple methods are based on  Kalman filter, particle filter, and their variants. Other
approaches are based on hidden Markov models. Most recent works have focused on different motion models, local navigation schemes, and machine learning methods.

\subsection{Human Trajectory Prediction}
Many motion and local navigation models have been proposed to model pedestrian trajectories.
These include model-based methods that describe pedestrian local movement through dynamics formulations, hand-crafted features or rules. Some of the commonly used models are based on social forces~\cite{7} and its variants~\cite{14,16}. Other techniques are based on reciprocal velocity obstacles~\cite{van2011reciprocal,2019061401,2019061803}, energy-based formulations~\cite{pellegrini2009you}, flocking rules~\cite{19}, continuum dynamics~\cite{8}, Gaussian processes~\cite{29}, and other techniques used to estimate the intent~\cite{25,2019061303,2019061802}.

The model-based methods have been used for trajectory estimation and prediction. However, their performance depends on the choice of underlying parameters and may vary based on the context and pedestrian behaviors.
Other set of techniques for pedestrian modeling are based on data-driven methods. These are driven by large number of real-world trajectories or related datasets or examples combined with some rules or learning methods.

Data-driven methods extract crowd movement characteristics from a large amount of real (video) data to predict individual trajectories.
Lerner et al.~\cite{6} predict the trajectory of each agent by searching the database for an example that closely matches the current prediction scene.
Later, \cite{33} improve this approach in terms of choosing the behavior and action of each agent.   However, their trajectory prediction method is strictly limited by the example database and unable to deal with new situations (e.g. with moving obstacles). Some combinations of data-driven and model-driven methods have also been proposed~\cite{kim2016interactive}. Our approach is also motivated by prior data-driven methods and we use the prior examples as `trajectory database structures'.
Inspired by the success of LSTM neural networks for different sequence prediction tasks, many researchers have used such networks for trajectory prediction~\cite{36,37,38,39,2019061801}.
However, LSTM neural networks usually deal with relatively short input streams~\cite{11} and may not work well in terms of long input stream information required for trajectory prediction~\cite{35,42}.
In addition, it is very difficult to extend these methods for new datasets. They usually need to retrain on the corresponding model, which can be time-consuming.

\subsection{Personality Models}


Personality has a great influence on human movement trajectory.
In the same scenario, different individuals with various personality traits often show different behaviors.
Some of these different personality traits are regarded as basic personality and they can be used to describe others. Trait theories \cite{3} are able to define these primary personality traits. The personality of one person can be described by the degrees of these primary traits he or she exhibits. The Eysenck 3-factor model \cite{43} is one of the most well-established trait theories. The primary personality traits of this model are Psychoticism, Extraversion, and Neuroticism (PEN).

In addition to the study of personality modeling, other researchers have explored the relationship between personality and movements \cite{3,4}.
Guy et al. \cite{3} describe the relationship between simulation parameters and personality descriptors based on the PEN personality model. Bera et al. \cite{4} present a real-time algorithm to automatically classify the dynamic behavior or personality of one pedestrian in a crowd video according to Personality Trait Theory.
In \cite{13}, crowds are classified as audiences and mobs. Considering personality is not enough to represent an impulsive mob agent, they introduce an emotion component that modulates agents' decision making processes superimposed on their personalities.
Durupinar et al. \cite{44} map personality trait to set of behaviors in the HiDAC (High-Density Autonomous Crowds) crowd simulation system. Using this method, users no longer need to do the tedious task of low-level parameter tuning. Besides, this approach is able to combine all these behaviors with different personality factors.

Based on the well-established PEN model, we are able to estimate human personality from crowd movement for individual trajectories prediction.
However, most of the existing studies focus on the relationship between parameters of specific crowd simulation algorithm and individual personality. Few of them focus on the quantitative relationship between individual movement trajectories and personality characteristics. Inspired by~\cite{4}, we further explore this relationship based on the movement characteristics of pedestrians, and use the extracted personality features in our probabilistic map.

\section{Prediction Using Probabilistic Feature Map}
In this section, we present our trajectory prediction method based on a probabilistic feature map. Our goal is to use a map representation that  fuses different factors, including the human personality. Moreover, these personality features are estimated by analyzing long-term historical trajectories of pedestrians, which are more stable and accurate for later trajectory prediction.

\begin{figure*}[htb] \begin{centering}
  \centering
  \includegraphics[width=18cm]{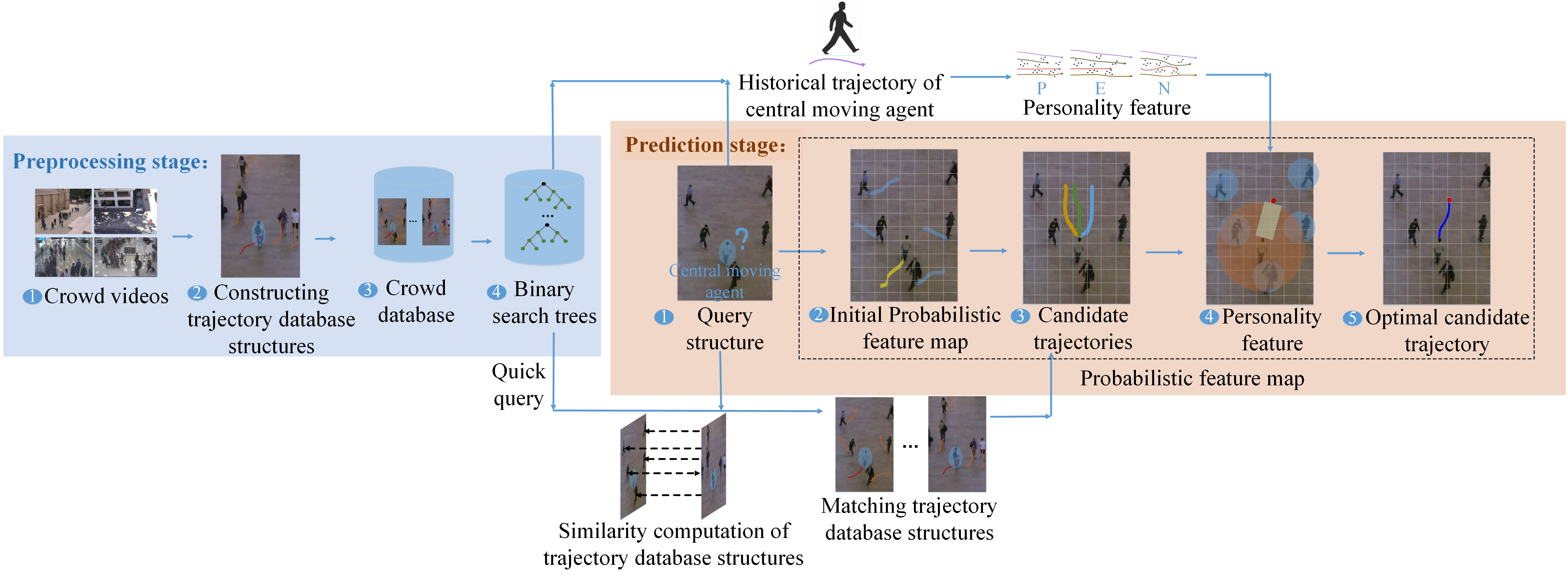}
  \centering
  \caption{
 {\bf Overview of Our Algorithm}. During the pre-processing stage, (1) and (2) pedestrian and crowd videos are processed into a lot of trajectory database structures, which correspond to real-world example data. (3) All the trajectory database structures are stored in the crowd database. (4) The trajectory database structures are classified according to their similarity and stored in self-balancing binary search trees. During the prediction stage, the following procedures are performed. (1)
A spatial query structure is computed based on surrounding scene information. (2) A probabilistic feature map is generated according to the query structure and the map is divided into multiple grids.
(3) Based on similarity computations on the trajectory database structures, several matchings are computed in the crowd database and used to compute candidate trajectories.
Based on these trajectories, we also predict the immediate goal of the central moving agent.
 (4) The weights associated with the grids are updated according to the personality features, which are calculated based on prior trajectories of the central moving agent.
 (5) The shortest path from the start point to the destination represents the predicted trajectory.
  }
  \label{fig:1}
  \end{centering}
\end{figure*}

An overview of our approach is given
in Figure \ref{fig:1}. Our approach is divided into two stages: pre-processing and prediction. During the pre-processing stage, we process a vast amount of real-world crowd videos into a large number of examples (referred as {\em trajectory database structure}) stored in a crowd database (Section \ref{Information unit}). During the prediction stage, we construct a probability feature map by adjusting its weights integrating several important influence factors (Section \ref{Probabilistic feature map}). The predicted trajectory for each pedestrian is obtained based on the probability feature map (Section \ref{Optimal trajectory prediction}).

\subsection{Trajectory Database Structure}
\label{Information unit}

\begin{figure}[htb] \begin{centering}
  \centering
  \includegraphics[width=8.3cm]{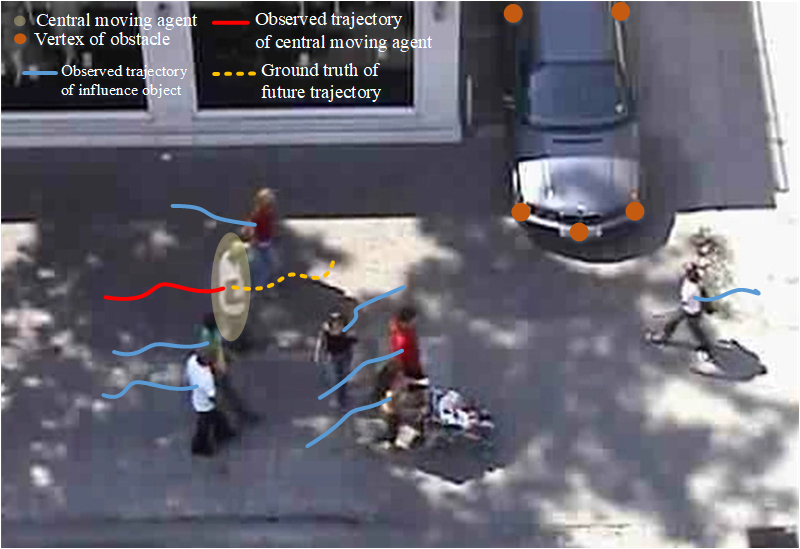}
  \centering
  \caption{
  Example of a trajectory database structure. The red line represents the historical trajectory of the central moving agent. The yellow dotted line is the real trajectory of the central moving agent in the future $p$ frames. The blue lines represent the trajectories of surrounding individuals, which will influence the future trajectory of the central moving agent. The obstacle is represented by a polygon and the vertexes of the polygon are marked.
  }
  \label{fig:2}
  \end{centering}
\end{figure}

We precompute a large number of trajectory database structures, which are extracted from real-world crowd movement trajectories.
 Trajectory database structure is designed for a single pedestrian, which corresponds to the {\em central moving agent}, and includes his or her surrounding scene information. In Figure \ref{fig:2}, we show the schematic diagram of the trajectory database structure.
This structure consists of three parts:
\begin{itemize}
\item The historical trajectory of the central moving agent.
\item The trajectories or locations of the influencing objects and obstacles around the central moving agent.
\item The actual trajectory of the central moving agent in the future.
\end{itemize}
The time length of a trajectory database structure is $f$ frames. The space covered by it corresponds to a rectangle of $2w$ pixels $\times$ $w$ pixels. The trajectory segment $T_k$ of an individual is represented as a set of ordered discrete points, denoted as $T_k = \{ {p_i} = ({x_i},{y_i})|i = 1,2,...,n_t\}$. $n_t$ indicates that the movement is distributed on $n_t$ frames. $({x_i},{y_i})$ represents the position of the individual in the $i$-th frame.
${T_{gt}} = \{ p_i^{gt} = (x_i^{gt},y_i^{gt})|i = 1,2,...,p\}$ denotes the actual movement trajectory of the central moving agent in the next $p$ frames.
An obstacle $O_k$ is represented by a polygon and is denoted as $O_k=\{v_i=(x_i,y_i)|i=1,2,...,n_v\}$. $v_i$ represents a vertex of the polygon and $n_v$ indicates the number of the vertices.
Trajectory database structure can be defined as in Equation (\ref{eq1000}):
\begin{equation}\label{eq1000}
\scalebox{1}
{$
{I_j} = \{ {T_k}|k = 1,2, \ldots ,m_1\} \cup \{ {O_i}|i = 1,2, \ldots ,m_2\} \cup {T_{gt}},
$}
\end{equation} 
where $I_{j}$ is a trajectory database structure, $m_1$ is the number of individuals contained in the trajectory database structure, and $m_2$ is the number of the obstacles in trajectory database structure.


\textbf{Query structure} is a special kind of trajectory database structure, which is formed  based on the latest movement states of individuals in the current prediction scene.
It only contains the observed trajectories or locations of the surrounding  individuals or obstacles and does not contain the trajectory of the central moving agent in the next $p$ frames. We compute query structure to predict the future trajectory.

\textbf{1. Fast Matching}

\begin{figure}[htb] \begin{centering}
  \centering
  \includegraphics[width=8.8cm]{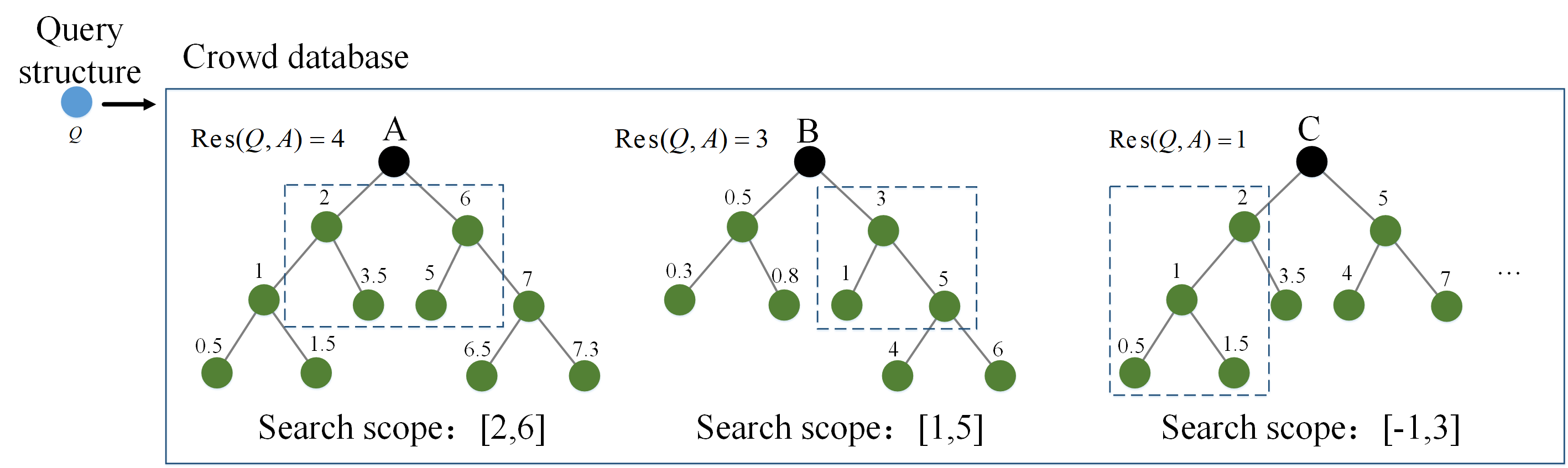}
  \centering
  \caption{
  Quickly finding matching trajectory database structures. The black circles are the representative trajectory database structures. The green circles are common trajectory database structures. The number next to the green circle represents the similarity value between the common trajectory database structure and the representative one.
  }
  \label{fig:3}
  \end{centering}
\end{figure}

Based on the similarity measurement defined in \cite{6}, we present a fast matching algorithm to find top-$k$ similar trajectory database structures to the query structure.
Figure \ref{fig:3} is the schematic of this.

We assume that the crowd database contains $N$ trajectory database structures. During the preprocessing stage, we choose $m$ representative trajectory database structures ($m \ll N$) that differ considerably from each other.
They represent different types of pedestrian movements.
The similarities between all common trajectory database structures and the $m$ representative ones are computed. Next, each representative trajectory database structure is taken as the root node of a self-balancing binary search tree, along with other similar common trajectory database structures.

During the prediction stage, we search the matching trajectory database structures from these trees. 
For each parent node, the similarity of all nodes in the left subtree is less than that of all nodes in the right subtree. If the similarity between the query structure $Q$ and the representative trajectory database structure $A$ is $\eta$, i.e. $Res(Q,A)=\eta$, we can only search the trajectory database structures of the corresponding binary tree that have similarities between $\eta-\delta$ and $\eta+\delta$, which reduces the search range. In this paper, we set $\delta=2$ considering the trade-off accuracy and efficiency. The search range of other self-balancing binary search trees with representative trajectory database structures is determined in the same way.

\subsection{Probabilistic Feature Map}
\label{Probabilistic feature map}

\label{3.2.1 Definition}

The probabilistic feature map is represented using a concrete query structure. The spatial range covered by this query structure is divided into grids. The weight of each grid records the possibility of the central moving agent passing through this grid during the future $p$ frames. The larger the weight of a grid is, the greater the possibility that the central moving agent passes through it. Through leveraging the weights of different grids, the probabilistic feature map can account for different factors that govern the trajectory prediction computation. In this paper, we consider several influence factors including surrounding pedestrians, static obstacles, candidate trajectories, destination, and personality features.

We use a matrix $W$ to represent the probabilistic feature map. It is an $m$-by-$n$ matrix and $w_{ij}$ is the weight of the grid in the $i$-th row and $j$-th column.
\begin{equation}\label{eq12}
W = \left[ {\begin{array}{*{20}{c}}
{{w_{11}}}&{{w_{12}}}& \cdots &{{w_{1n}}}\\
{{w_{21}}}&{{w_{22}}}& \cdots &{{w_{2n}}}\\
 \cdots & \cdots & \ddots & \cdots \\
{{w_{m1}}}&{{w_{m2}}}& \cdots &{{w_{mn}}}
\end{array}} \right] = ({w_{ij}}) \in {\mathbb{R}^{m \times n}}
\end{equation} 

\textbf{1. Weight Updating}

Each influencing factor changes the weight of each grid in the probabilistic feature map and is denoted as $\varphi $. The weights are updated in Equation (\ref{eq13}).
\begin{equation}\label{eq13}
w_{ij}=w_{ij}^{o} + \varphi_{in}(i,j) + \varphi_{o}(i,j) + \varphi_{c}(i,j) + \varphi_d(i,j) + \varphi_p(i,j),
\end{equation} 
where $w_{ij}^{o}$ is the initial weight of the grid in the $i$-th row and $j$-th column and $w_{ij}$ is the final weight of it.
$\varphi_{in}$, $\varphi_{o}$, $\varphi_{c}$, $\varphi_d$, and $\varphi_p$ represent the effect of surrounding individuals, obstacles, candidate trajectories, the predicted destination, and personality factors on the weight of the grid, respectively.

\textbf{1) Influence of surrounding pedestrians or static obstacles}

In query structure ${Q}$, $m$ is the central moving agent and $n$ represents another pedestrian, who has influence on the movement trajectory of $m$.
The influence is calculated according to the influence function ${{\rm Imp(m,n,t)}}$ based on the distance and moving speed of individuals \cite{6}.
${{\rm Imp(m,n,t)}}$ represents the influence of the pedestrian $n$ on the pedestrian $m$ at time $t$. If $n$ is in front of $m$, Equation (\ref{eq1}) is used to calculate the influence. If $n$ is behind $m$, Equation (\ref{eq2}) is used to calculate the influence.
\begin{equation}\label{eq1}
{\mathop{\rm Im}\nolimits} {\rm{p}}(m,n,t) = \exp (\frac{{ - 0.5{d^2}}}{{2/v}}) \cdot \exp (\frac{{ - 0.5{d^2}}}{v}),
\end{equation} 
\begin{equation}\label{eq2}
{\mathop{\rm Im}\nolimits} {\rm{p}}(m,n,t) = \exp (\frac{{ - 0.5{d^2}}}{{2/v}}) \cdot \exp (\frac{{ - 0.5{d^2}}}{{1/2v}}),
\end{equation} 
where $v$ is moving speed of the individual $m$ and $d$ is the distance from $m$ to $n$, respectively.

According to the influence values, we update the weights of the map. $\varphi_{in}(i,j)$ is defined using the following Equation:
\begin{equation}\label{eq1001}
\varphi_{in}(i,j)=-1 \cdot \sum_{n=1}^{ n_{in}}{ \sum_{t=1}^{f}{ \textbf{1}_{ij}[x,y] \cdot 0.1^{f-t} \cdot 3 \cdot \rm Imp(m,n,t)}},
\end{equation} 
where $n_{in}$ is the number of the individuals who have influence on the agent $m$ and $(x,y)$ is the coordinate of the individual $n$ at the $t$-th frame. $\textbf{1}_{ij}[x,y]$ is defined as follows:
\begin{equation}\label{eq1002}
\textbf{1}_{ij}[x,y] =\left\{
\begin{array}{rcl}
1       &      &   \text{if} \ (x,y) \ \text{is}  \ \text{in} \ \text{the} \ \text{grid} \ (i,j),\\
0    &      &  \text{otherwise}.\\
\end{array} \right.
\end{equation} 
$\varphi_{o}(i,j)$ is defined as following:
\begin{equation}\label{eq1003}
\varphi_{o}(i,j) =\left\{
\begin{array}{rcl}
- \infty      &      &   \text{if} \ (i,j) \ \text{is}  \ \text{in}  \ \text{an} \ \text{obstacle},\\
0    &      & \rm otherwise.\\
\end{array} \right.
\end{equation} 

\textbf{2) The influence of candidate trajectories and destination}

Using the fast query method proposed in Section~\ref{Information unit}, $k$ candidate trajectory database structures are computed. We can
get $k$ candidate trajectories from them. In our current implementation, we use $k=3$.  Meanwhile, the weights of the grids that these candidate trajectories pass through are updated.
\begin{equation}\label{eq1004}\scriptsize
\varphi_{c}(i,j)= \alpha \cdot \sum_{t=1}^{p} {\textbf{1}_{ij}[x_1^t,y_1^t]} + \beta \cdot \sum_{t=1}^{p} {\textbf{1}_{ij}[x_2^t,y_2^t]}
+ \gamma \cdot \sum_{t=1}^{p} {\textbf{1}_{ij}[x_3^t,y_3^t]},
\end{equation} 
where $\alpha$, $\beta$, and $\gamma$ are the weights of three candidate trajectories. $\alpha=30$, $\beta=15$, and $\gamma=10$. $(x_1^t,y_1^t)$, $(x_2^t,y_2^t)$, and $(x_3^t,y_3^t)$ are the points on the three candidate trajectories at the $t$-th frame.

Based on the candidate trajectories and known trajectories of the central moving agent, the destination location of the central moving agent can be predicted using Equation (\ref{eq14}).
In addition, the weight of the grid in which the destination locates is changed based on Equation (\ref{eq1005}).
\begin{equation}\label{eq14}
D = w{s_1} \cdot {D_1} + w{s_2} \cdot {D_2} + w{s_3} \cdot {D_3} + wcs \cdot {D_{cs}},
\end{equation} 
where $D$ is the predicted destination. $D_1$, $D_2$, and $D_3$ are the destinations of candidate trajectories. $ws_1$, $ws_2$, and $ws_3$ are their weights, respectively. $D_{cs}$ is the symmetric starting point of the known trajectory for the agent (assuming that the time length of the known trajectory is equal to that of the predicted trajectory) and $wcs$ is the weight of $D_{cs}$.
\begin{equation}\label{eq1005}
\varphi_{d}(i,j)= \varepsilon \cdot \textbf{1}_{ij}[x,y]
\end{equation} 
where $\varepsilon \to +\infty $ and $(x,y)$ is the predicted destination.

\textbf{3) Personality features}

We use the well-established PEN personality model \cite{43} to describe human personalities. This model consists of three independent factors of personality: Psychoticism, Extraversion, and Neuroticism. The pedestrian with high P-factor usually takes the most direct path and doesn't mind the distance to the surrounding pedestrians. The pedestrian with high E-factor usually walks faster than ordinary people. The pedestrian with high N-factor keeps his or her distance from surrounding pedestrians ignoring the distance to the destination~\cite{3}.

Based on the known long-term historical trajectory of the central moving agent $i$ of a query structure, we calculate a three-dimensional personality feature vector $(l,v,d)$  representing the personality features of the agent $i$. $l$ is the linearity of the trajectories of the agent $i$, $v$ is the average speed of it, and $d$ is the average minimum distance of the agent $i$ to surrounding individuals.



\begin{figure}[htbp]
\centering
\begin{tabular}{ccc}
 \includegraphics[width=2.6cm]{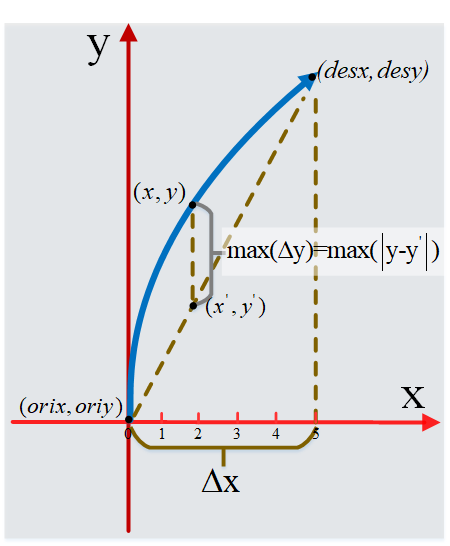} &  \includegraphics[width=2.6cm]{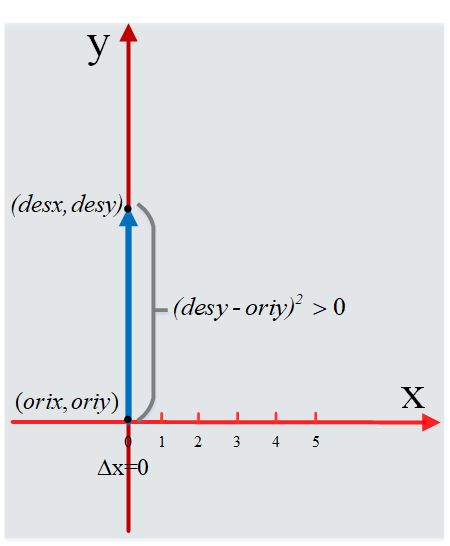} & \includegraphics[width=2.6cm]{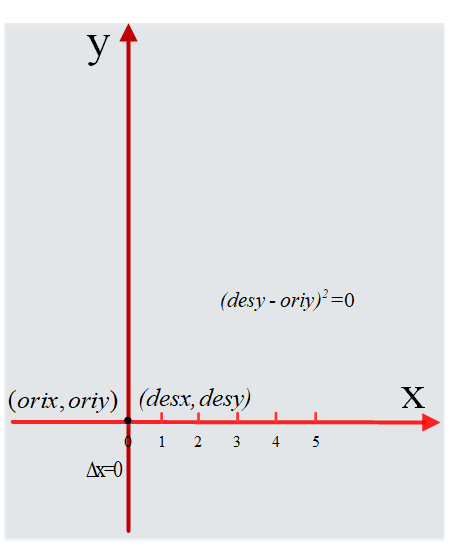}\\
 (a)   & (b)  & (c)  \\
\end{tabular}
\caption{
Calculating the linearity of trajectory. The blue line is the real trajectory of the individual in 5 frames. $(x,y)$ is the coordinate of the point in the real trajectory. $(x^{'},y^{'})$ is the coordinate of the point in the linear trajectory from the beginning $(orix, oriy)$ to the end $(desx, desy)$. $\Delta{x} = |desx - orix|$, $\Delta y = |y-y^{'}|$, and $x^{'}=x$. (a) $\Delta {{x}} \ne 0 $. (b) The individual moves in the positive direction along the $y$ axis (i.e. $\Delta{x} = 0$ and $(desy - oriy)^2>0$). (c) The individual is stationary in 5 frames
(i.e. $\Delta{x} = 0$ and $(desy - oriy)^2=0$).
}
\label{fig:7}
\end{figure}

Inspired by \cite{Hu2014The}, we present the definition of the linearity.
In Figure \ref{fig:7}, we show the schematic diagram for calculating the linearity of trajectory
assuming that the time length of the trajectory database structure is 5 frames.
The linearity of a point on the real trajectory segment is defined as following:

\begin{equation}\label{eq1010}\scriptsize
Linearity(x,y)=\left\{\begin{matrix}
{\frac{{{ \left| {y-y\mathop{{}}\nolimits^{{'}}} \right| }}}{{\Delta x}} \times 100\text{\%}} & \text{if }\Delta x \ne 0\\
0 & \text{if }\Delta x=0 \text{ and } (desy-oriy)^{2}  > 0\\
-1 & \text{if }\Delta x=0 \text{ and } (desy-oriy)^{2}  = 0
\end{matrix}\right.
\end{equation} 
where $(x,y)$ is the coordinate of one point on the real trajectory segment $T$, $(x^{'},y^{'})$ is the coordinate of one point on the linear trajectory from the starting point$(orix, oriy)$ to the end $(desx, desy)$ of $T$, $x^{'}=x$,
$\Delta x = |desx - orix|$, and $Linearity(x,y)$ is the linearity of the point $(x,y)$ on $T$.

\begin{equation}\label{eq1011}
Linearity(T)= max \ Linearity(x,y),
\end{equation} 
where $(x,y)$ is the coordinate of any point on $T$ and $Linearity(T)$ is the linearity of the trajectory segment $T$.
The historical trajectory of the agent $i$ is divided into several trajectory segments of $f$ frames.
 $l$ is the average value of linearity over all the fragments, which is define as following.
\begin{equation}\label{eq120190715}
{l=\frac{{1}}{{N_l}} \cdot {\mathop{ \sum }\limits_{{i=1}}^{{\mathop{{N_l}}}}{Linearity \left( T_i \right) }}},
\end{equation} 
 where $N_l$ is the number of the trajectory segments.
 The higher the value of $l$, the higher the P-factor of his or her personality.

 The E-factor of individual personality is measured by $v$. $v$ is the average speed of all historical trajectories of the agent $i$, which
 is defined in Equation \ref{eq12019071502}.
 We measure the N-factor of one individual personality by $d$. $d$ is the average minimum distance of the agent $i$ to surrounding
individuals, which is defined in Equation \ref{eq12019071503}. The higher the value of $d$, the higher the N-factor of his or her personality.

\begin{equation}\label{eq12019071502}
{v=\frac{{{\mathop{ \sum }\limits_{{i=2}}^{{t}}{dis \left( \mathop{{P}}\nolimits_{{i}},P\mathop{{}}\nolimits_{{i-1}} \right) }}}}{{t-1}}}
\end{equation} 
where $dis(P_i,P_{i-1})$ denotes the distance between the position at the time $t$ and the position at time $t+1$ of an individual.

\begin{equation}\label{eq12019071503}
{d=avg \left( \mathop{{min}}\limits_{{j \in D\text{ } \wedge \text{ }j \neq i}} \left( d_{ij}   \right)  \right) }
\end{equation} 
where $D$ contains all the individuals of the query structure. $d_{ij}$ is the distance between the individual $i$ and the individual $j$.

The weights of the grids are updated according to the personality factors as following:
\begin{equation}\label{eq1006}
\varphi_{p}(i,j)= \varphi_{l}(i,j) + \varphi_{v}(i,j) + \varphi_{d}(i,j),
\end{equation} 
\begin{equation}\label{eq1007}
\varphi_{l}(i,j) =\left\{
\begin{array}{rcl}
\mu      &      &   \text{if} \ Linearity(i,j) \leq l, \\
0    &      & \text{otherwise},\\
\end{array} \right.
\end{equation} 
\begin{equation}\label{eq1008}
\varphi_{v}(i,j) =\left\{
\begin{array}{rcl}
\nu      &      &    \text{if} \ (i-orix)^2+(j-oriy)^2 \leq v \cdot p, \\
0    &      & \text{otherwise},\\
\end{array} \right.
\end{equation} 
\begin{equation}\label{eq1009}
\varphi_{d}(i,j) =\left\{
\begin{array}{rcl}
\eta      &      &   \text{if} \ dis(i,j) \le d,  \\
0    &      & \text{otherwise},\\
\end{array} \right.
\end{equation} 
where $(orix,oriy)$ is the starting point of our predicted trajectory, $dis(i,j)$ is the distance from the grid $(i,j)$ to the central moving agent, $\mu=20$, $\nu=10$, $\eta=-10$, and $p$ is the time of predicted trajectory.

\begin{figure*}[t]
\centering
\begin{tabular}{cccc}
 \includegraphics[width=3.9cm]{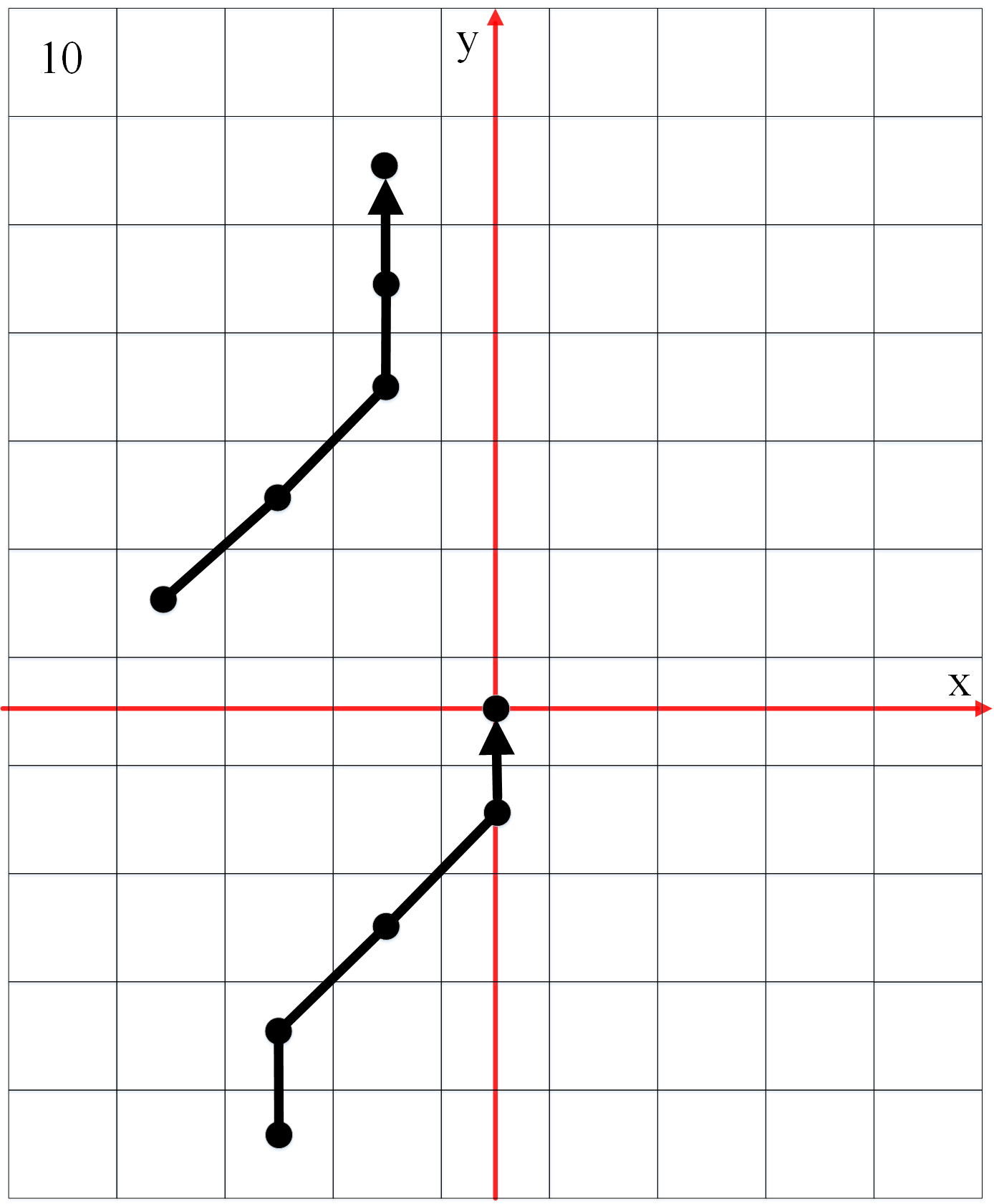} &  \includegraphics[width=3.9cm]{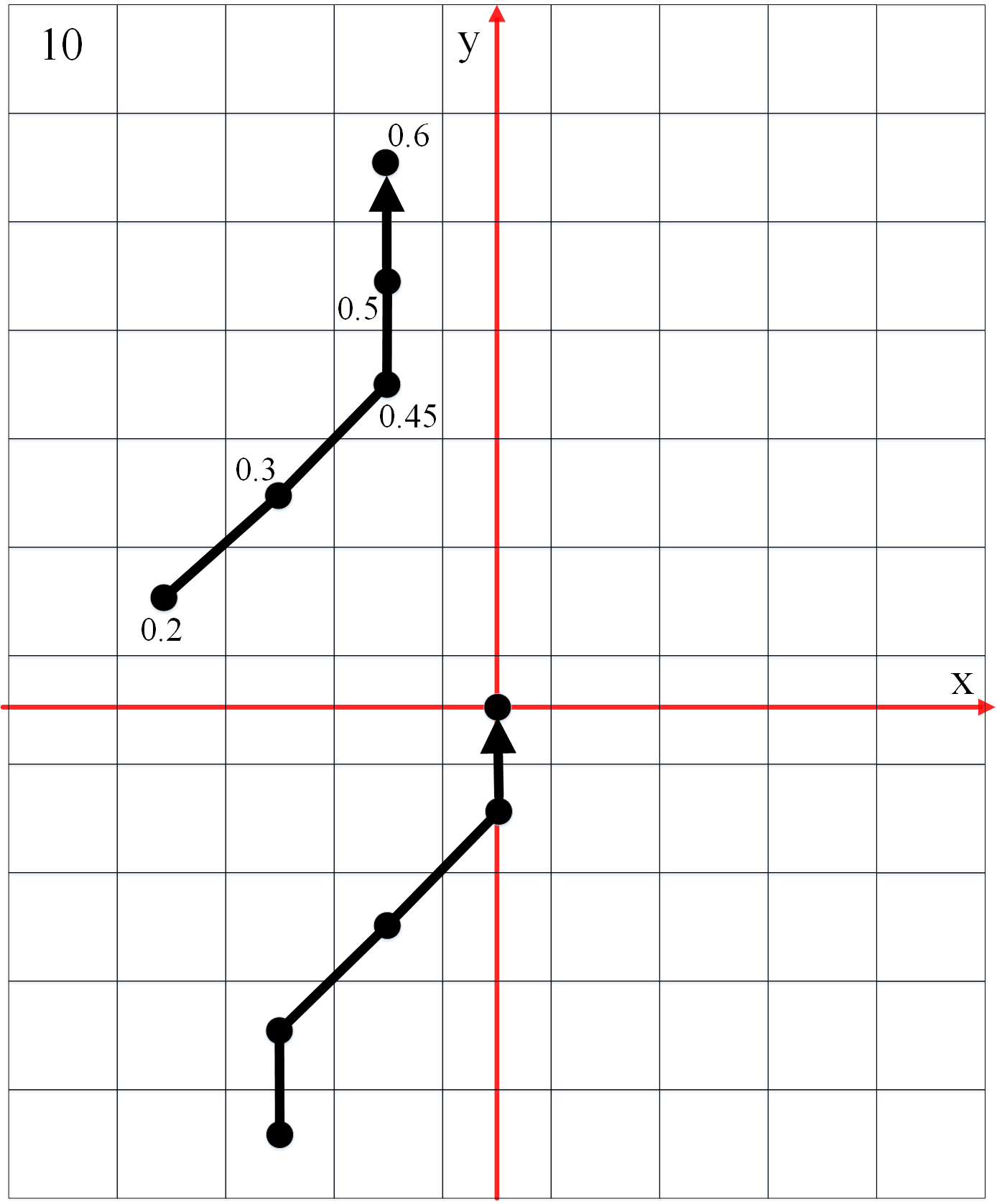}&\includegraphics[width=3.9cm]{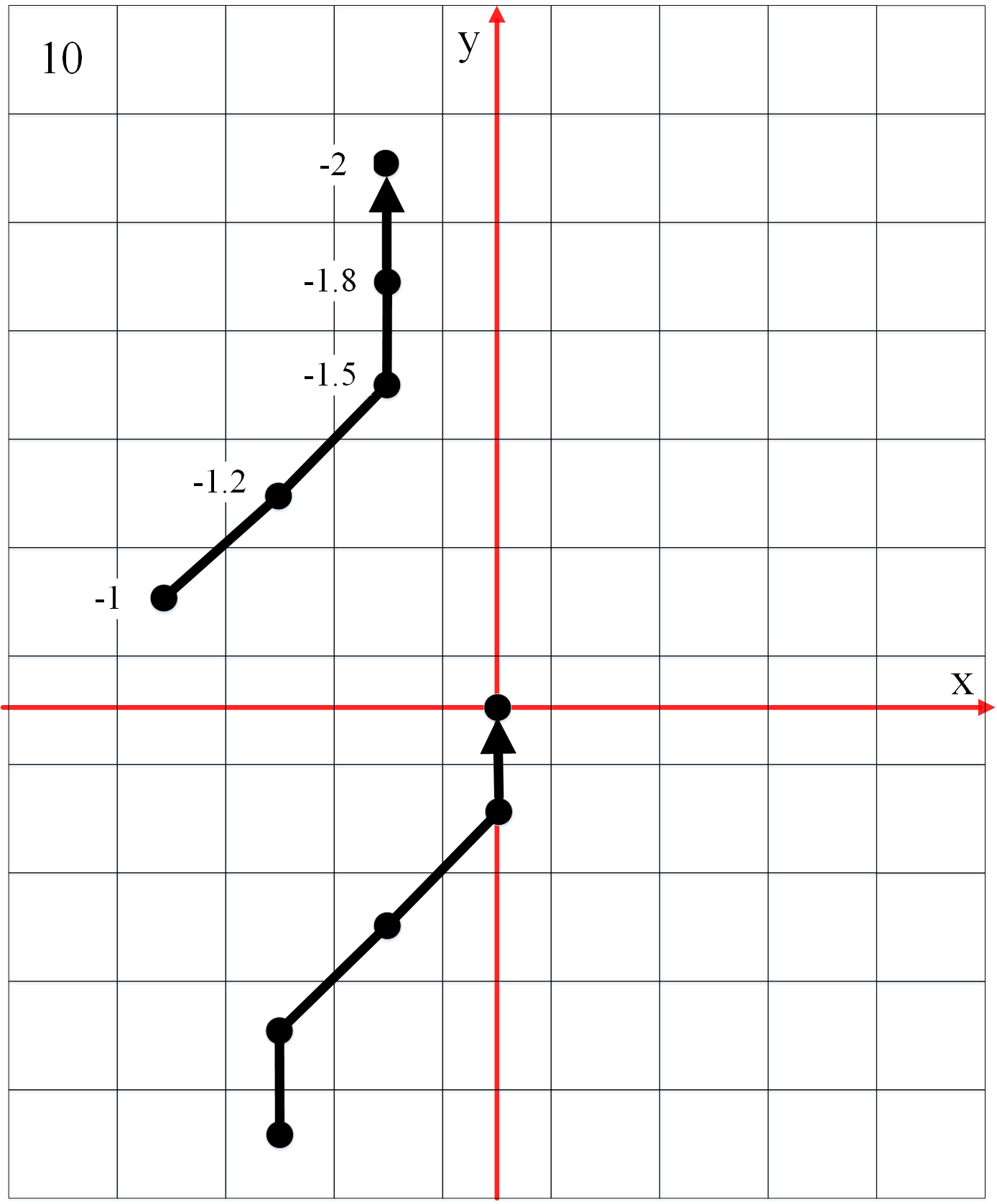}&  \includegraphics[width=3.9cm]{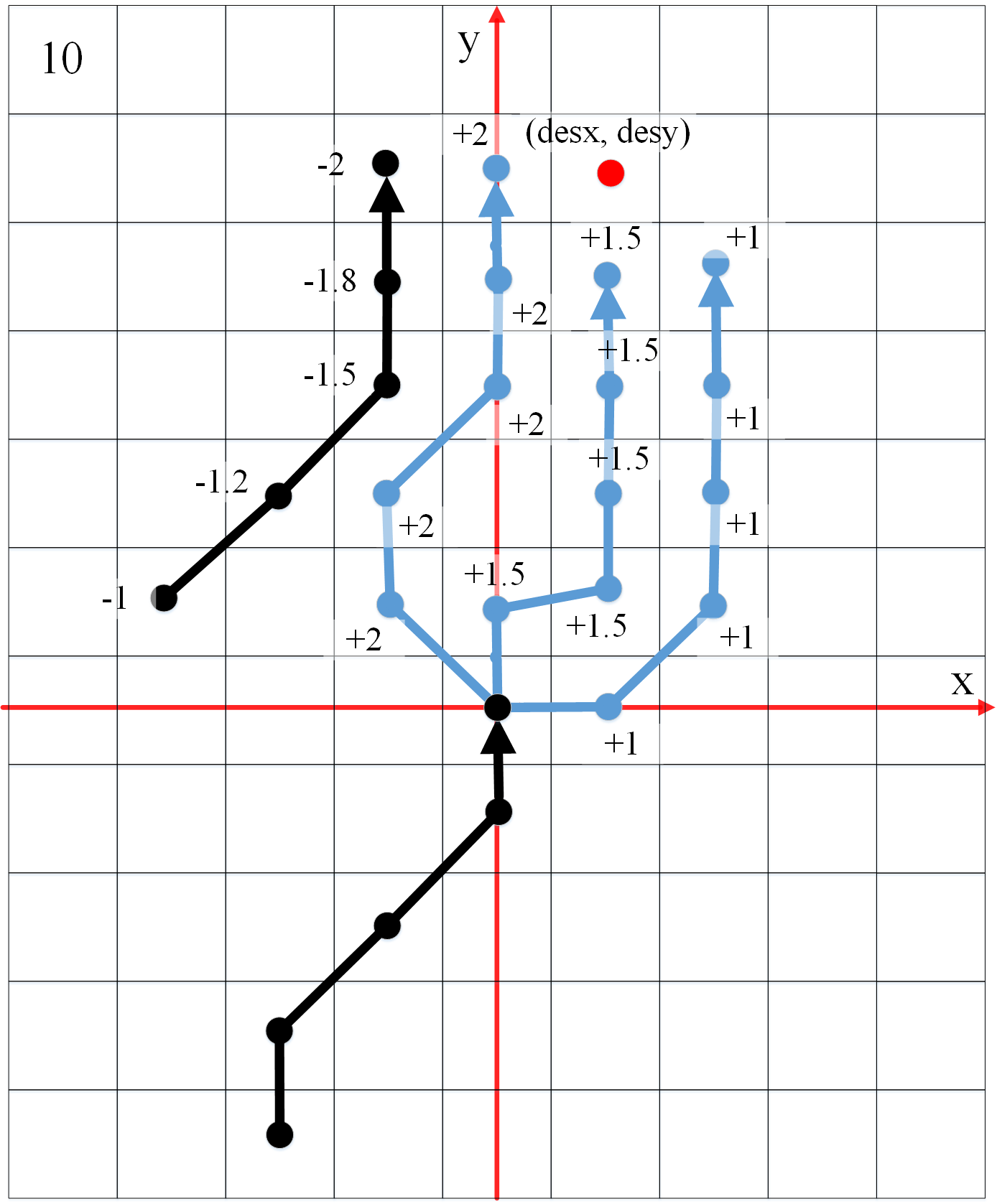}\\
 (a)   & (b) &(c)& (d)\\
  \includegraphics[width=3.9cm]{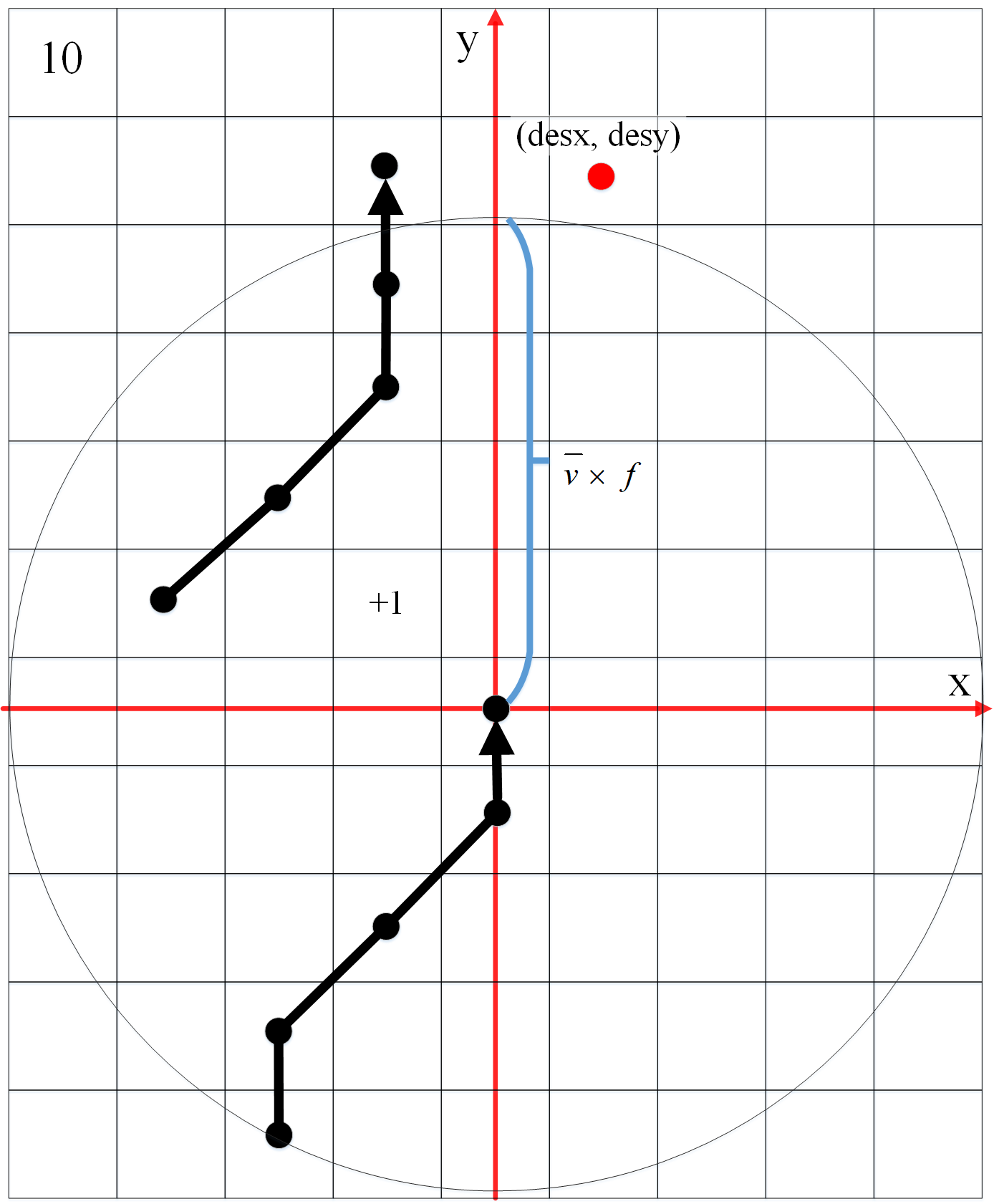} &  \includegraphics[width=3.9cm]{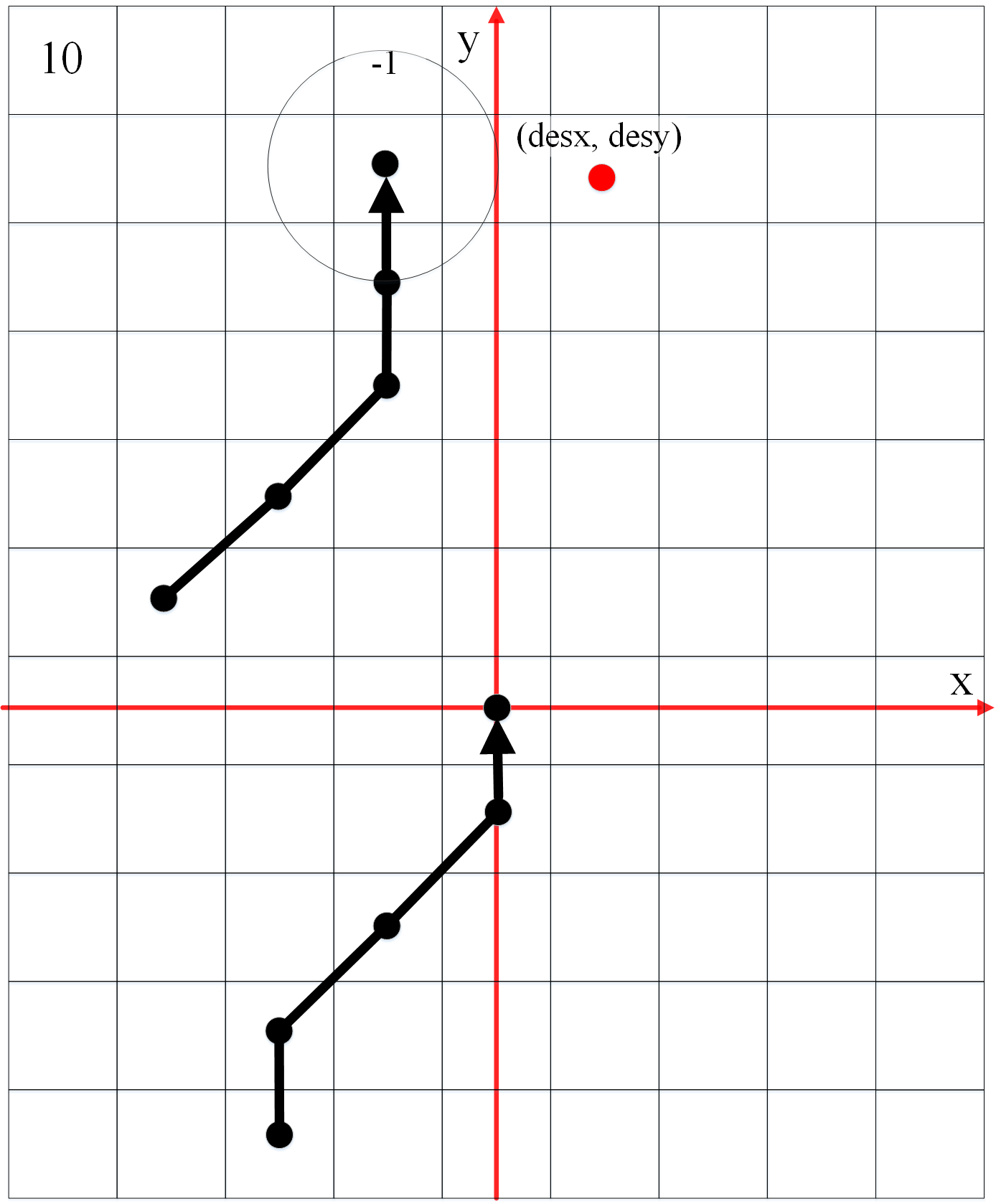}&\includegraphics[width=3.9cm]{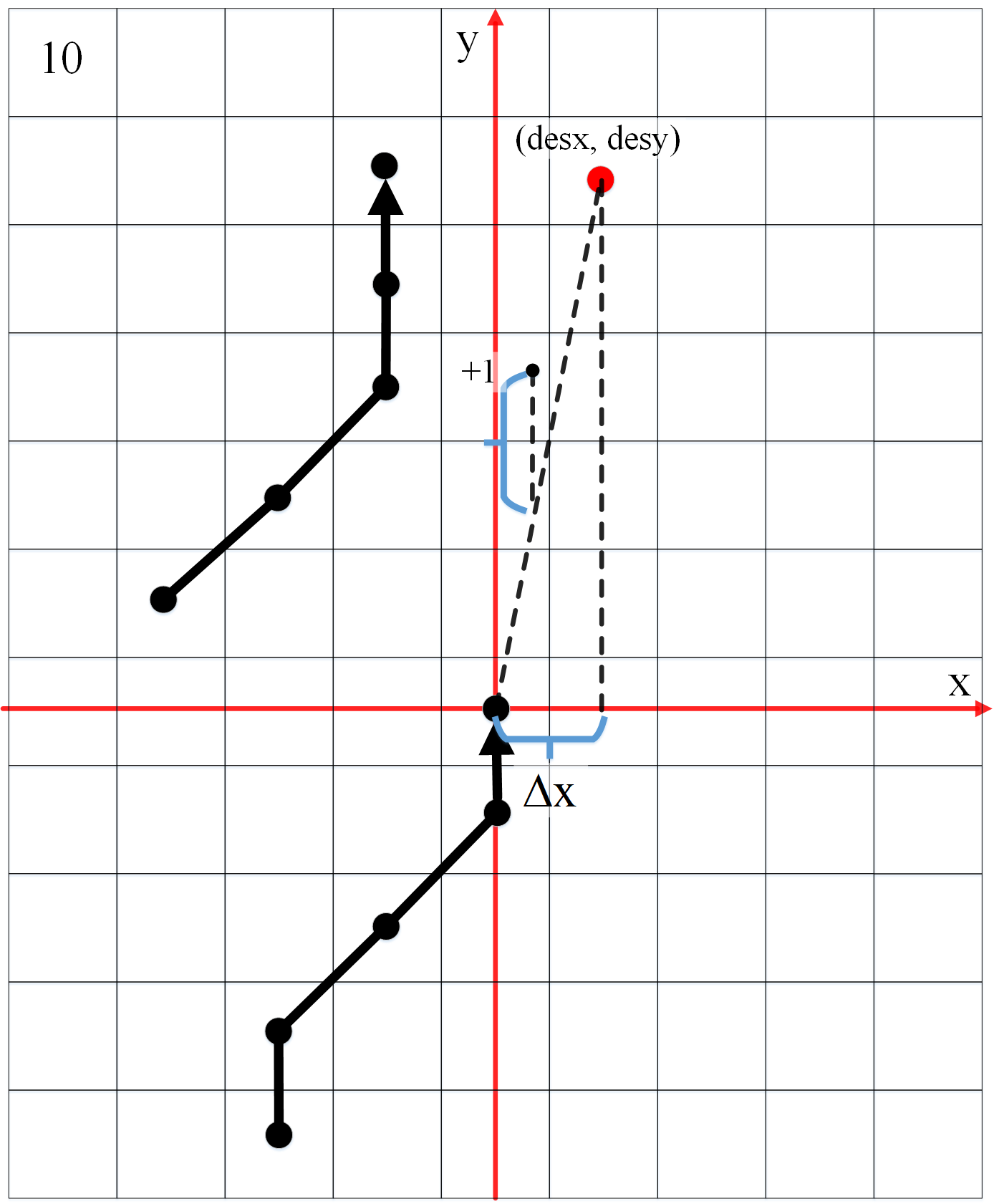}&  \includegraphics[width=3.9cm]{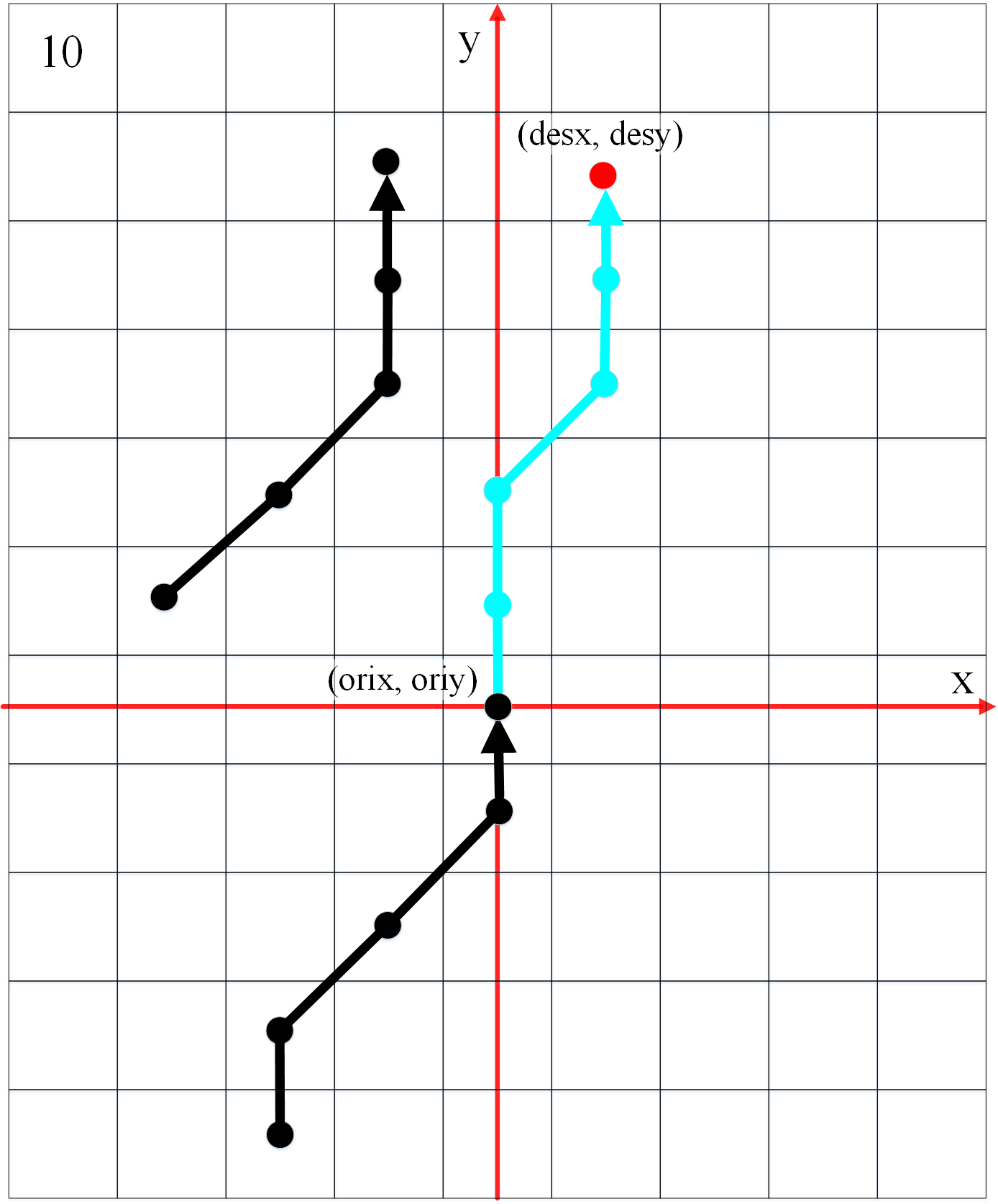}\\
 (e)   & (f)&(g)& (h)\\
\end{tabular}
\caption{
Weight updating of the probabilistic feature map. (a) The probabilistic feature map is divided into a lot of grids. Two black curves represent two trajectory segments. The points on the curve represent the grids that the trajectory passes through. (b) The influence of the surrounding individuals on the central moving agent is calculated. (c) According to the influence values in (b), the values of the weights in the corresponding grids are reduced. (d) Three candidate trajectories in crowd database are found and the possibility values of corresponding grids are increased. In (e), (f), and (g) the weights of probabilistic feature map are updated according to personality features. (h) The final prediction trajectory is obtained by estimating the shortest path.
}
\label{fig:55}
\end{figure*}

\begin{table*}[htbp]
\setlength{\belowcaptionskip}{0.3cm}
\centering
\caption{List of parameter values used in our model.}
\begin{tabular}{p{6.5cm}<{\raggedright}|m{4cm}<{\raggedright}|m{4cm}<{\raggedright}}
\hline
Parameters                                                                                       & Sign                   & Values                                                                              \\ \hline
Size of trajectory database structure                                                                         & $2w \times w$                       & $7m\times3.5m$  \cite{33}                                                                    \\ \hline
Time length of trajectory database structure                                                                      & $f$                       & 3.2s(80 frames) \cite{50}                                                              \\ \hline
Time length of prediction trajectory                                                                & $p$                       & \begin{tabular}[c]{@{}l@{}}3.2s(80 frames);\\ 4.8s (120 frames) \cite{50} \end{tabular} \\ \hline
The size of a grid in probabilistic feature map                                           & $l\_g \times w\_g$                       & $0.2m\times0.2m$                                                                             \\ \hline
Initial value of probabilistic feature map                                                & $P{V_{initial}}$                       & 5                                                                                   \\ \hline
Number of matching trajectory database structures found                                                       & $Nu{m_{{\rm{matching}}\_iu}}$                       & 3                                                                                   \\ \hline
Possibility value of candidate trajectories updating the probabilistic feature map                          & $P{V_{c1}}$,$P{V_{c2}}$, and $P{V_{c3}}$                       & +30;+15;+10                                                                         \\ \hline
Weights of the predicted destination in Equation \ref{eq14}                                              & $ws_1$, $ws_2$, $ws_3$, and $wcs$ & 0.3, 0.15, 0.10, and 0.80                                                             \\ \hline
Possibility value of average moving speed updating the probabilistic feature map                            & $P{V_{ave\_speed}}$                       & +10                                                                                 \\ \hline
Possibility value of minimum distance to the surrounding individuals updating the probabilistic feature map & $P{V_{\min \_dis}}$                        & -10
\\ \hline
Possibility value of linearity updating the probabilistic feature map                                       & $P{V_{lin}}$                       & +20                                                                                 \\ \hline
Number of representative trajectory database structures                                                       & $Nu{m_{rep\_iu}}$                       & 28                                                                                  \\ \hline
\end{tabular}
\label{table:1}
\end{table*}

\subsection{Trajectory Prediction}
\label{Optimal trajectory prediction}

 Based on the probabilistic feature map constructed in Section \ref{Probabilistic feature map}, we compute a shortest path from the starting point $(orix,oriy)$ to the destination $(desx,desy)$ based on the Dijkstra algorithm. This shortest path must satisfy two conditions at the same time: the possibility values of the passing grids are large and the length of path is small. The shortest path is the final prediction trajectory.

In Figure \ref{fig:55}, we give an example to show how to update the weights of the probabilistic feature map. The weights are updated according to following steps: (a) In Figure \ref{fig:55}a, the space depicted by the probabilistic feature map is divided into grids
with same size. Each grid stores a weight representing the
possibility of the predicted pedestrian passing through, assuming that the initial probability value of each grid is 10. (b) The influence of the surrounding individuals on the central moving agent is calculated as we show in Figure \ref{fig:55}b. (c) According to the influence values of (b), the values of the weights in the corresponding grids are reduced. (d) Three candidate trajectories in crowd database are found and the possibility values of the corresponding grids are increased in which the candidate trajectories are located. $(desx, desy)$ is the destination we predict. (e) The average speed of the central moving agent multiplied by the predicted time determines the possible range of the movement in the future $p$ frames, and the possible values of the corresponding grids are increased. (f) According to long-term historical trajectory of the central moving agent, the average minimum distance away from the surrounding individuals is obtained and the probability values of the corresponding grids are decreased. (g) The possibility values of the corresponding grids are updated according to the linearity. (h) The final prediction trajectory is obtained by estimating the shortest path.

\section{Implementation and Performance}

We have implemented our model on a PC with a quadcore 2.50 GHz CPU, 16GB memory, and an Nvidia GeForce GTX 1080 Ti graphics card. We evaluated our method on two publicly available datasets: ETH \cite{48} and UCY \cite{49}.

\subsection{Experiment Settings}
\label{Experiment setting}

Similar to prior methods~\cite{50}, we report the prediction error using two different metrics.
\begin{itemize}
\item Average Displacement Error (ADE): Average L2 distance between ground truth and our prediction over all predicted time steps.
\item Final Displacement Error (FDE): The distance between the predicted final destination and the true final destination at the end of the prediction period.
\end{itemize}
Furthermore, we
compare the results using the following baselines:

 {\em Linear model}: A linear regressor that estimates linear parameters by minimizing the least square error.

 {\em Sim-1}: We report the results of a simplified version of our model which finds the most similar trajectory database structure to query structure in the crowd database. The real future trajectory of the trajectory database structure is taken as the prediction result without using probabilistic feature map.

   {\em Sim-$k$}: This method is also a simplified version of our model which finds the top-$k$ matching trajectory database structures. The final prediction result is obtained by weighted averaging of these $k$ candidate results.

{\em S-LSTM}: The method proposed by Alahi et al.~\cite{2}

{\em SGAN}: The method proposed by Gupta et al.~\cite{50}

{\em GLMP}: The method proposed by Bera et al.~\cite{32}

To compare our method with other methods impartially, a summary of the parameter values used is depicted in Table \ref{table:1}.

\begin{table*}[htbp]
\setlength{\belowcaptionskip}{0.2cm}
\centering
\caption{Quantitative results of different methods across the datasets. The ETH dataset is split into two sets (ETH and HOTEL). The UCY dataset has 3-components: UNIV, ZARA-01, and ZARA-02. We report two error metrics ADE and FDE for different prediction times ($3.2$s and $4.8$s) in meters.
Our method outperforms the state-of-the-art methods for most cases, as shown in bold faces, though it is comparable to SGAN in some cases. AVG represents the average performance over all scenarios. }
\begin{tabular}{|c|c|c|c|c|c|c|c|}
\hline
{Metric} & {Dataset} & {Linear} & {Sim-1} & {Sim-$k$} & {S-LSTM} & {SGAN}               & Our model \\ \hline
\multirow{5}{*}{ADE}    & ETH                      & 0.84/1.33               & 0.60/0.96              & 0.60/0.91       & 0.73 / 1.09       &  0.61 / 0.81 &  \textbf{0.51}/\textbf{0.73} \\
                        & HOTEL                    & 0.35/0.39               & 0.37/0.58              & 0.35/0.55      &0.49 / 0.79        &  0.48 / 0.72 &  \textbf{0.20}/\textbf{0.24} \\
                        & UNIV                     & 0.56/0.82               & 0.69/1.11              & 0.53/0.85       &0.41 / 0.67       &  \textbf{0.36} / \textbf{0.60} &  0.47/0.70 \\
                        & ZARA01                   & 0.41/0.62               & 0.55/0.89              & 0.45/0.74       &0.27 / 0.47       &  \textbf{0.21} / \textbf{0.34} &  0.50/0.64 \\
                        & ZARA02                   & 0.53/0.77               & 0.45/0.71              & 0.42/0.66       &0.33 / 0.56       &  \textbf{0.27} / 0.42 & 0.30/\textbf{0.41} \\ \hline
AVG                     &                          & 0.54/0.79               & 0.53/0.85              & 0.47/0.74       &  0.45 / 0.72       &  \textbf{0.39} / 0.58 & 0.40/\textbf{0.54} \\ \hline
\multirow{5}{*}{FDE}    & ETH                      & 1.60/2.94               & 1.21/1.91              & 1.20/1.80        & 1.48 / 2.35      &  1.22 / 1.52 &  \textbf{1.04}/\textbf{1.44} \\
                        & HOTEL                    & 0.60/0.72               & 0.73/1.16              & 0.69/1.09        & 1.01 / 1.76      &  0.95 / 1.61 & \textbf{0.31}/\textbf{0.43} \\
                        & UNIV                     & 1.01/1.59               & 1.44/2.34              & 1.11/1.82     & 0.84 / 1.40         & \textbf{0.75} / \textbf{1.26}  & 0.89/1.26 \\
                        & ZARA01                   & 0.74/1.21               & 1.16/1.87              & 0.95/1.56      &  0.56 / 1.00        &  \textbf{0.42} / \textbf{0.69}  & 0.90/1.16 \\
                        & ZARA02                   & 0.95/1.48               & 0.94/1.45              & 0.88/1.35       &  0.70 / 1.17       &  0.54 / 0.84 & \textbf{0.51}/\textbf{0.78} \\ \hline
AVG                     &                          & 0.98/1.59               & 1.11/1.74             & 0.97/1.52      &  0.91 / 1.54        &  0.78 / 1.18 &  \textbf{0.73}/\textbf{1.02} \\ \hline
\end{tabular}
\label{table:2}
\end{table*}

\subsection{Comparison with Other Models}

We compare our method using two metrics, ADE and FDE (lower numbers are better), against different baselines in Table \ref{table:2}. Our method is especially good for long-terms predictions. The prediction result for a time period of  $4.8$s are better than those for $3.2$s. The Linear model is only capable of modeling straight or constant speed paths; therefore, this kind of method cannot generate accurate prediction results. The Sim-1 model finds only one matching trajectory database structure and uses that to compute predicted trajectory. The Sim-$k$ model considers $k$ (in this paper $k=3$) matching trajectory database structures to calculate prediction results. The Sim-1 model performs worse than the Sim-$k$ model.
Our method chooses multiple candidate trajectories based on a probabilistic feature map method.
It can integrate complicated influence factors on trajectory prediction intuitively and efficiently by adjusting the weights stored in different grids.
We also consider the individual personality factors which are closely related to the movement trajectories. Therefore, our model outperforms the Sim-1 and Sim-$k$ models.

In most cases, our model performs better than the SGAN and S-LSTM models because each predicted sample can account for multiple possible future trajectories. The SGAN model predicts social plausible futures by training adversarial against a recurrent discriminator and encouraging diverse predictions with a novel variety loss. However, the SGAN and S-LSTM models only consider the short-term trajectories of individuals and the interaction between individuals. In addition, our model also considers individual long-term trajectories that integrate individual personality and other influence factors.
On average, our algorithm performs better than SCAN and S-LSTM. Although our prediction result might be a little different from ground truth data, it can represent one of many possible future predictions.

We also compare our model with the GLMP model \cite{32} in Table \ref{table:2019061201}. The Bayesian Inference method is used to learn pedestrian local and global movement patterns from real pedestrian trajectory data for the GLMP model. We present a trajectory prediction algorithm
for pedestrians based on a personality-aware probabilistic feature map. This map is computed according to the prior trajectory of each pedestrian and
the surrounding scene information. Furthermore, we estimate the personality characteristics of each pedestrian and
use them to improve the prediction. Therefore, our model outperforms the GLMP model.

\begin{figure}[t]
\centering
\begin{tabular}{cc}
 \includegraphics[width=4cm]{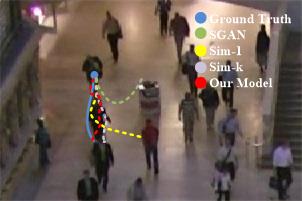} &  \includegraphics[width=4cm]{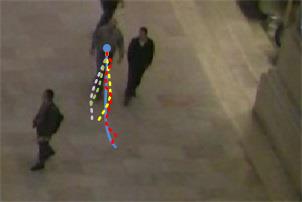}  \\
   (a) & (b)   \\
  \includegraphics[width=4cm]{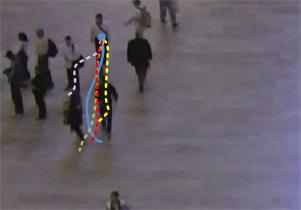} &  \includegraphics[width=4cm]{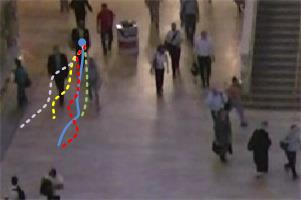}  \\
   (c) & (d)  \\
  \includegraphics[width=4cm]{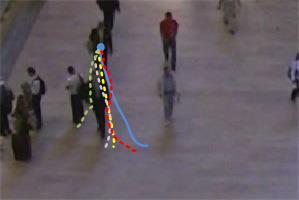} &  \includegraphics[width=4cm]{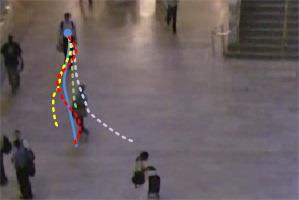}  \\
   (e) & (f)  \\
\end{tabular}
\caption{We highlight the performance of different trajectory prediction algorithms in the  trajectories scene from  Grand Central Train Station in New York city. The solid blue lines represent the ground truth trajectories. The green, yellow, and purple dashed lines represent the prediction results of the SGAN, Sim-1, and Sim-$k$ models, respectively. The red dotted lines represent our prediction results, which is more accurate. We highlight the benefits of our approach in six configurations from (a) to (f).}
\label{fig:10}
\end{figure}

We qualitatively evaluate the performance of different methods on real scenes where individuals interact with each other.
In the trajectories scene from Grand Central Train Station in New York city \cite{Zhou2011Random} (Figure \ref{fig:10}),
we illustrate the prediction results of our model, Sim-1, Sim-$k$, and SGAN. More results can be seen in the
supplementary video. Each person is denoted by a cylinder. We marked trajectories of all the persons.
Our model is better at predicting individual trajectories than other methods.

\begin{table}[htbp]
\setlength{\belowcaptionskip}{0.3cm}
\centering
\caption{Comparison with the GLMP model \cite{32} across the datasets \cite{6977426}. We highlight the results
for short-term prediction (1 sec) and long term prediction (5 sec).
We evaluate the accuracy of long and short term predictions. If the mean error between the
prediction result and the ground-truth value at a time instance is less than 0.8 meter in ground space coordinates, the prediction is considered
to be ``successful'' \cite{32}.
We notice that our model results in higher accuracy than the GLMP model.}
\begin{tabular}{|c|c|c|c|c|}
\hline
\multirow{2}{*}{Dataset} & \multicolumn{2}{c|}{GLMP} & \multicolumn{2}{c|}{Our model} \\ \cline{2-5}
                         & 1 sec       & 5 sec       & 1 sec          & 5 sec         \\ \hline
NDLS-1                   & 60.2\%      & 51.2\%      & 71.4\%         & 61.9\%        \\ \hline
IITF-1                   & 71.2\%      & 50.5\%      & 82.0\%         & 57.3\%        \\ \hline
IITF-3                   & 68.4\%      & 45.7\%      & 79.2\%         & 51.1\%        \\ \hline
IITF-5                   & 64.6\%      & 40.0\%      & 74.1\%         & 48.6\%        \\ \hline
\end{tabular}
\label{table:2019061201}
\end{table}

\subsection{Ablation Study}
\label{Ablation study}

In this section, we evaluate our model through the following ablation experiments. We discuss the effect of the following influence factors on our performance. 

\textbf{Effect of Different Numbers of Candidate Trajectories }

We compare the performance of our model with different numbers of candidate trajectories. Table \ref{table:3} shows the ADE and FDE of our prediction results with 1, 2, 3, 4, and 5 candidate trajectories.
The performance of our model can be improved to some extent by increasing the number of candidate trajectories.
However, when the number of candidate trajectories is greater than 3, the performance cannot be improved. The reason is that the redundant candidate trajectories cannot match the current prediction scenarios very well and they have little effect on trajectory prediction. In addition, they may also increase the complexity of our algorithm.
Assuming that the crowd database contains $n$ trajectory database structures, the time complexity of finding the top-$k$ matching trajectory database structures in $n$ trajectory database structures is $O(n*logk)$. As a result, $k=3$ offers the best balance in terms of accuracy and efficiency.
\begin{table}[htbp]
\setlength{\belowcaptionskip}{8pt}
\centering
\caption{The ADE and FDE of our prediction results with different candidate trajectories. We highlight the minimum values of ADE and FDE,  which indicates that we can obtain the prediction results with the smallest error when the number of candidate trajectories is 3.}

\begin{tabular}{|c|c|c|c|c|c|}
\hline
\multirow{2}{*}{Metric} & \multicolumn{5}{c|}{The number of candidate trajectories} \\ \cline{2-6}
                        & 1         & 2         & 3         & 4         & 5         \\ \hline
ADE                     & 0.4326          &   0.4039        &    \textbf{0.3920}       &  0.3967         &  0.3968         \\ \hline
FDE                     &   0.7881        &     0.7281      &     \textbf{0.7012}     &    0.7290       &      0.7291     \\ \hline
\end{tabular}
\label{table:3}
\end{table}

\begin{table}[htbp]
\setlength{\belowcaptionskip}{5pt}
\centering
\caption{
The ADE of our prediction results with and without integrating human personality in different scenes. We obtain better results for ADE with the personality factors.
}
\begin{tabular}{|c|l|l|}
\hline
Dataset & \multicolumn{1}{c|}{ADE(w/o)} & \multicolumn{1}{c|}{ADE(w)} \\ \hline
ETH     &      0.7208/1.0458                         &     \textbf{0.7155}/\textbf{1.0020}                                           \\ \hline
HOTEL   &    0.3976/0.5201     &\textbf{0.3820}/\textbf{0.5081}                             \\ \hline
UNIV    &   0.4646/\textbf{0.7597}    &   \textbf{0.4497}/0.7605                             \\ \hline
ZARA01  &   0.4719/0.6823  & \textbf{0.4701}/\textbf{0.6813}                             \\ \hline
ZARA02  &     \textbf{0.5617}/\textbf{0.8132}   &  0.5648/0.8139                                \\ \hline
AVG     &       0.5233/0.7642 & \textbf{0.5164}/\textbf{0.7533}                            \\ \hline
\end{tabular}
\label{table:4}
\end{table}

\textbf{Benefits of Human Personality Modeling}

We illustrate the relationship between the accuracy of our model and the personality factors. We compare the ADE of our prediction results with and without integrating human personality in Table \ref{table:4}. After integrating the personality factors, the accuracy of our prediction results for ADE is improved by 1.4\% on the average.
In most cases, personality features can improve predicted results. In some special cases, an individual's behavior may not reflect his or her personality features.
For example, a calm person who is anxious about train will increase his or her moving speed, which is different from his or her general behaviors.

\textbf{Impact of the weights of destination prediction on performance}

\begin{figure}[htbp]
\centering
\begin{tabular}{c}
 \includegraphics[width=8.5cm]{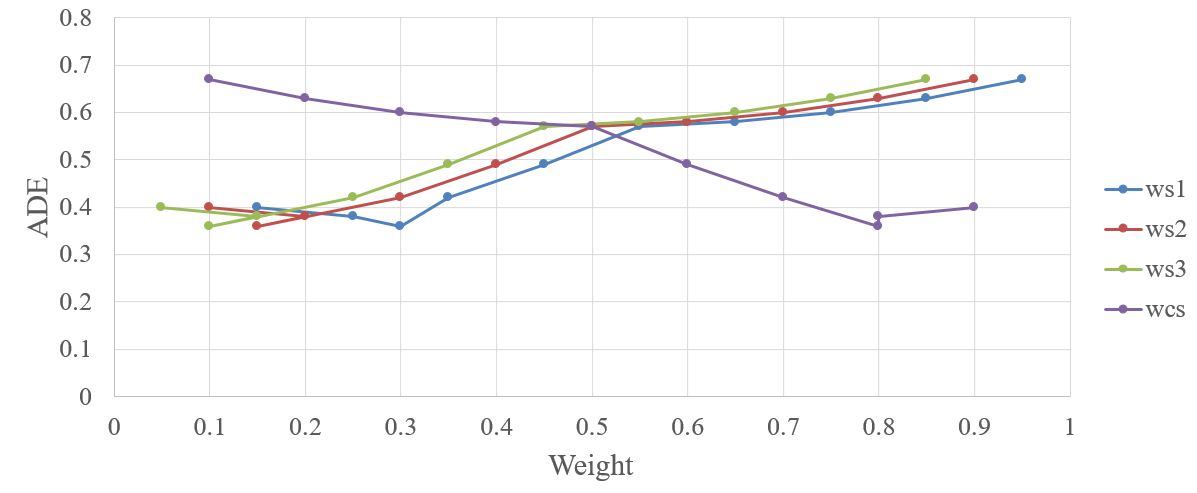}  \\
 (a)     \\
   \includegraphics[width=8.5cm]{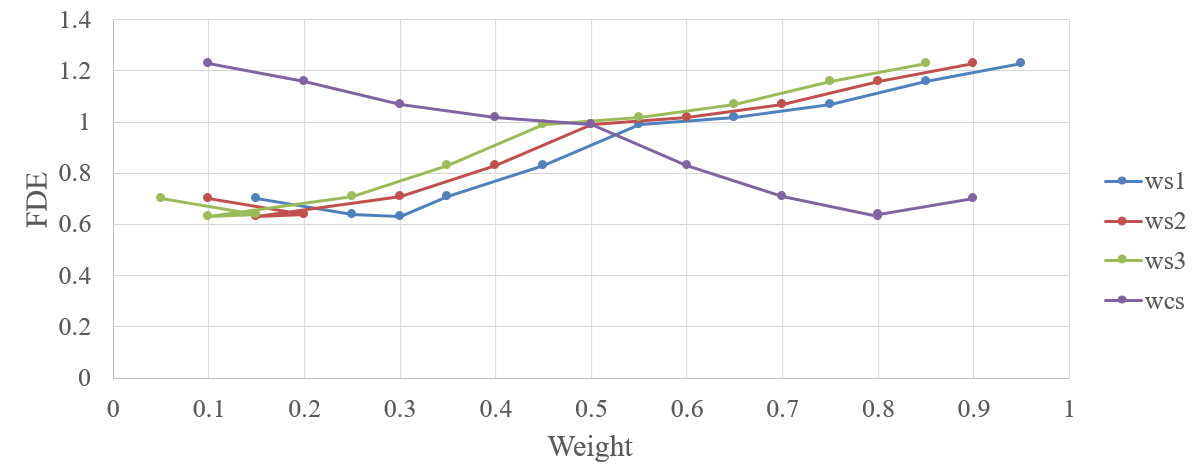} \\
  (b)  \\
\end{tabular}
\caption{
The ADE and FDE of our prediction results with different values of these weights ($ws_1$, $ws_2$, $ws_3$, and $wcs$).
(a) The $x$ and $y$ axes correspond to the weights and ADE. (b) The $x$ and $y$ axes correspond to the weights and FDE.
}
\label{fig:11}
\end{figure}

The weights ($ws_1$, $ws_2$, $ws_3$, and $wcs$) for destination prediction in Equation (\ref{eq14}) are important to compute the final trajectory prediction results.
We evaluate the relationship between the accuracy of trajectory prediction and the values of the parameters ($ws_1$, $ws_2$, $ws_3$, and $wcs$). In Figure \ref{fig:11}, we show the ADE and FDE of our prediction results with different values of these parameters. When ADE and FDE are at their minimum, $ws_1$, $ws_2$, $ws_3$, and $wcs$ are 0.30, 0.15, 0.10, and 0.80, respectively, which are the optimal values of these parameters. At this time, $wcs$ is much higher than the weights of the candidate trajectories. Therefore, the predicted destination is mainly related to the known trajectory of the agent in the query structure.

\begin{figure}[htb] \begin{centering}
  \centering
  \includegraphics[width=8cm]{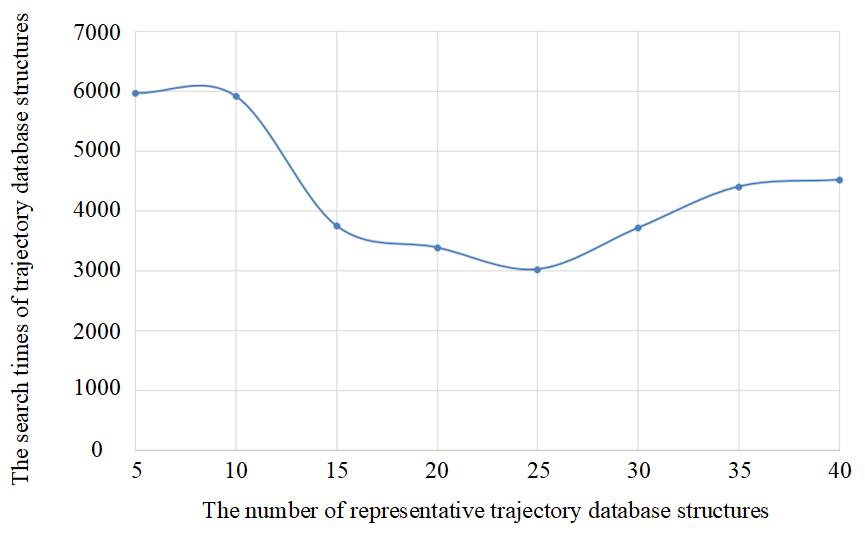}
  \centering
  \caption{The $x$-axis and $y$-axis correspond to the number of representative trajectory database structures and the search times of trajectory database structures.}
  \label{fig:12}
  \end{centering}
\end{figure}

\subsection{Analysis of query efficiency}

Figure \ref{fig:12} illustrates the relationship between the search times and the number of representative trajectory database structures when we want to find matching trajectory database structures in our crowd database. When this number increases from 5 to 25, the search times of common trajectory database structures decrease. The increase of this number means that the average height of the self-balancing binary search tree decreases. Therefore, the search times of common trajectory database structures decrease. As the number of representative trajectory database structures increases (from 25 to 40), the number of self-balancing binary search trees that need to be searched also increases. Therefore, the search times of common trajectory database structures increase. The search times of trajectory database structures by our algorithm are far less than the total number of trajectory database structures (30536). We can find the matching trajectory database structures by traversing an average of 3024 common trajectory database structures (about 10\% of the total number). Our method can greatly reduce the scope of traversing the crowd database.

\section{Conclusions and Limitations}

We present a personality-based individual trajectory prediction model for probabilistic feature map. Our approach accounts for different factors that influence the  trajectory of a pedestrian in terms of map computation.  The predicted trajectory corresponds to the shortest path in the map. Based on the classic PEN personality model, we quantify the relationship between individual trajectories and personality features and improve the accuracy of our prediction method. We have evaluated the performance on standard benchmarks and $5-9$\% improvement is observed in accuracy.

Our approach has some limitations. It is a data-driven method, whose accuracy depends on the specific crowd videos used to pre-compute the trajectory database structures. If the new pedestrian video is very different from the pre-computed database, our accuracy may degrade. There is no simple way to compute the optimal weights for different scenarios and it may be useful to explore learning techniques. The prediction accuracy can change if the scene density changes or there is a fast moving obstacle.

\ifCLASSOPTIONcaptionsoff
  \newpage
\fi



%
%
%

\bibliographystyle{IEEEtran}
\bibliography{bare_jrnl}

\begin{thebibliography}{10}
\providecommand{\url}[1]{#1}
\csname url@samestyle\endcsname
\providecommand{\newblock}{\relax}
\providecommand{\bibinfo}[2]{#2}
\providecommand{\BIBentrySTDinterwordspacing}{\spaceskip=0pt\relax}
\providecommand{\BIBentryALTinterwordstretchfactor}{4}
\providecommand{\BIBentryALTinterwordspacing}{\spaceskip=\fontdimen2\font plus
\BIBentryALTinterwordstretchfactor\fontdimen3\font minus
  \fontdimen4\font\relax}
\providecommand{\BIBforeignlanguage}[2]{{%
\expandafter\ifx\csname l@#1\endcsname\relax
\typeout{** WARNING: IEEEtran.bst: No hyphenation pattern has been}%
\typeout{** loaded for the language `#1'. Using the pattern for}%
\typeout{** the default language instead.}%
\else
\language=\csname l@#1\endcsname
\fi
#2}}
\providecommand{\BIBdecl}{\relax}
\BIBdecl

\bibitem{2019061301}
N.~{Takemura}, Y.~{Nakamura}, Y.~{Matsumoto}, and H.~{Ishiguro}, ``A
  path-planning method for human-tracking agents based on long-term
  prediction,'' \emph{IEEE Transactions on Systems, Man, and Cybernetics, Part
  C (Applications and Reviews)}, vol.~42, no.~6, pp. 1543--1554, Nov 2012.

\bibitem{2019061302}
Y.~{Zhou}, H.~{Hu}, Y.~{Liu}, S.~{Lin}, and Z.~{Ding}, ``A real-time and fully
  distributed approach to motion planning for multirobot systems,'' \emph{IEEE
  Transactions on Systems, Man, and Cybernetics: Systems}, pp. 1--15, 2018.

\bibitem{1}
F.~Bartoli, G.~Lisanti, L.~Ballan, and A.~Del~Bimbo, ``Context-aware trajectory
  prediction,'' in \emph{International Conference on Pattern Recognition},
  2018.

\bibitem{3}
S.~J. Guy, S.~Kim, M.~C. Lin, and D.~Manocha, ``Simulating heterogeneous crowd
  behaviors using personality trait theory,'' in \emph{ACM
  Siggraph/eurographics Symposium on Computer Animation}, 2011.

\bibitem{4}
A.~Bera, T.~Randhavane, and D.~Manocha, ``Aggressive, tense or shy? identifying
  personality traits from crowd videos,'' in \emph{Twenty-sixth International
  Joint Conference on Artificial Intelligence}, 2017.

\bibitem{13}
F.~Durupinar, U.~Gudukbay, A.~Aman, and N.~I. Badler, ``Psychological
  parameters for crowd simulation: from audiences to mobs,'' \emph{IEEE
  Transactions on Visualization and Computer Graphics}, vol.~22, no.~9, pp.
  2145--2159, 2016.

\bibitem{5}
S.~Hang, J.~Zhu, Y.~Dong, and Z.~Bo, ``Forecast the plausible paths in crowd
  scenes,'' in \emph{Twenty-sixth International Joint Conference on Artificial
  Intelligence}, 2017.

\bibitem{7}
D.~Helbing and P.~Molnar, ``Social force model for pedestrian dynamics,''
  \emph{Physical Review E}, vol.~51, no.~5, p. 4282, 1995.

\bibitem{8}
A.~Treuille, S.~Cooper, Popovi, and Zoran, ``Continuum crowds,'' \emph{ACM
  Transactions on Graphics}, vol.~25, no.~3, pp. 1160--1168, 2008.

\bibitem{32}
A.~Bera, S.~Kim, T.~Randhavane, S.~Pratapa, and D.~Manocha, ``\protect{GLMP}-
  realtime pedestrian path prediction using global and local movement
  patterns,'' in \emph{IEEE International Conference on Robotics \&
  Automation}, 2016.

\bibitem{36}
F.~Altch¨¦ and A.~de~La~Fortelle, ``An \protect{LSTM} network for highway
  trajectory prediction,'' in \emph{IEEE International Conference on
  Intelligent Transportation Systems}, 10 2017, pp. 353--359.

\bibitem{38}
S.~Haddad, M.~Wu, W.~He, and S.~K. Lam, ``Situation-aware pedestrian trajectory
  prediction with spatio-temporal attention model,'' in \emph{Computer Vision
  Winter Workshop}, 01 2019.

\bibitem{11}
X.~L. Shao, D.~Ma, Y.~Liu, and Y.~Quan, ``Short-term forecast of stock price of
  multi-branch \protect{LSTM} based on k-means,'' in \emph{International
  Conference on Systems \& Informatics}, 2018.

\bibitem{48}
S.~Pellegrini, A.~Ess, and L.~Van~Gool, ``Improving data association by joint
  modeling of pedestrian trajectories and groupings,'' in \emph{Proceedings of
  the 11th European Conference on Computer Vision}, 2010.

\bibitem{49}
L.~Leal-Taixe, M.~Fenzi, A.~Kuznetsova, B.~Rosenhahn, and S.~Savarese,
  ``Learning an image-based motion context for multiple people tracking,'' in
  \emph{IEEE Conference on Computer Vision \& Pattern Recognition}, 2014.

\bibitem{14}
A.~Alahi, V.~Ramanathan, and F.~F. Li, ``Socially-aware large-scale crowd
  forecasting,'' in \emph{IEEE Conference on Computer Vision \& Pattern
  Recognition}, 2014.

\bibitem{16}
R.~Emonet, J.~Varadarajan, and J.~M. Odobez, ``Extracting and locating temporal
  motifs in video scenes using a hierarchical non parametric bayesian model,''
  in \emph{IEEE Conference on Computer Vision \& Pattern Recognition}, 2011.

\bibitem{van2011reciprocal}
J.~Van Den~Berg, S.~J. Guy, M.~Lin, and D.~Manocha, ``Reciprocal n-body
  collision avoidance,'' in \emph{Robotics research}, 2011.

\bibitem{2019061401}
M.~Xu, X.~Xie, P.~Lv, J.~Niu, H.~Wang, C.~Li, R.~Zhu, Z.~Deng, and B.~Zhou,
  ``Crowd behavior simulation with emotional contagion in unexpected
  multi-hazard situations,'' \emph{IEEE Transactions on Systems, Man, and
  Cybernetics: Systems}, 02 2019.

\bibitem{2019061803}
Y.~Ma, D.~Manocha, and W.~Wang, ``Auto{RVO}: Local navigation with dynamic
  constraints in dense heterogeneous traffic,'' in \emph{International
  Conference on Control Systems and Computer Science}, 04 2018.

\bibitem{pellegrini2009you}
S.~Pellegrini, A.~Ess, K.~Schindler, and L.~Van~Gool, ``You'll never walk
  alone: Modeling social behavior for multi-target tracking,'' in \emph{IEEE
  International Conference on Computer Vision}, 2009.

\bibitem{19}
C.~W. Reynolds, ``Flocks, herds and schools: A distributed behavioral model,''
  \emph{ACM Siggraph Computer Graphics}, vol.~21, no.~4, pp. 25--34, 1987.

\bibitem{29}
J.~M. Wang, D.~J. Fleet, and H.~Aaron, ``Gaussian process dynamical models for
  human motion,'' \emph{IEEE Transactions on Pattern Analysis \& Machine
  Intelligence}, vol.~30, no.~2, pp. 283--298, 2007.

\bibitem{25}
S.~Lefevre, C.~Laugier, and J.~Ibaez-Guzman, ``Exploiting map information for
  driver intention estimation at road intersections,'' in \emph{Intelligent
  Vehicles Symposium}, 2011.

\bibitem{2019061303}
C.~{Anagnostopoulos} and S.~{Hadjiefthymiades}, ``Intelligent trajectory
  classification for improved movement prediction,'' \emph{IEEE Transactions on
  Systems, Man, and Cybernetics: Systems}, vol.~44, no.~10, pp. 1301--1314, Oct
  2014.

\bibitem{2019061802}
A.~Bera, T.~Randhavane, E.~Kubin, A.~Wang, D.~Manocha, and K.~Gray,
  ``Classifying group emotions for socially-aware autonomous vehicle
  navigation,'' in \emph{IEEE Computer Society Conference on Computer Vision
  and Pattern Recognition Workshops}, 05 2018, pp. 1039--1047.

\bibitem{6}
A.~Lerner, Y.~Chrysanthou, and D.~Lischinski, ``Crowds by example,'' in
  \emph{Computer Graphics Forum}, 2007.

\bibitem{33}
A.~Lerner, E.~Fitusi, Y.~Chrysanthou, and D.~Cohen-Or, ``Fitting behaviors to
  pedestrian simulations,'' in \emph{ACM Siggraph/eurographics Symposium on
  Computer Animation}, 2009.

\bibitem{kim2016interactive}
S.~Kim, A.~Bera, A.~Best, R.~Chabra, and D.~Manocha, ``Interactive and adaptive
  data-driven crowd simulation,'' in \emph{IEEE Virtual Reality}, 2016.

\bibitem{37}
B.~D. Kim, C.~M. Kang, S.~H. Lee, H.~Chae, J.~Kim, C.~C. Chung, and J.~W. Choi,
  ``Probabilistic vehicle trajectory prediction over occupancy grid map via
  recurrent neural network,'' \emph{Intelligent Transportation Systems
  Conference}, 2017.

\bibitem{39}
F.~Yang, H.~Saikia, and C.~Peters, ``Who are my neighbors?: A perception model
  for selecting neighbors of pedestrians in crowds,'' in \emph{Proceedings of
  the 18th International Conference on Intelligent Virtual Agents}, 2018, pp.
  269--274.

\bibitem{2019061801}
T.~Fernando, S.~Denman, S.~Sridharan, and C.~Fookes, ``Soft + hardwired
  attention: An {LSTM} framework for human trajectory prediction and abnormal
  event detection,'' \emph{Neural Networks}, vol. 108, 02 2017.

\bibitem{35}
N.~Lee, W.~Choi, P.~Vernaza, C.~Choy, P.~H.~S.~Torr, and M.~Chandraker,
  ``{DESIRE}: Distant future prediction in dynamic scenes with interacting
  agents,'' in \emph{Proceedings of the IEEE Conference on Computer Vision and
  Pattern Recognition}, 07 2017, pp. 2165--2174.

\bibitem{42}
Z.~Li, D.~He, F.~Tian, W.~Chen, T.~Qin, L.~Wang, and T.~Liu, ``Towards
  binary-valued gates for robust {LSTM} training,'' in \emph{Proceedings of the
  35th International Conference on Machine Learning}, vol.~80, 10--15 Jul 2018,
  pp. 2995--3004.

\bibitem{43}
H.~Eysenck and M.~Eysenck, ``Personality and individual differences: A natural
  science approach,'' \emph{Premuzic}, vol.~2, pp. 343--363, 1985.

\bibitem{44}
F.~Durupinar, N.~Pelechano, J.~M. Allbeck, U.~Gudukbay, and N.~I. Badler, ``How
  the ocean personality model affects the perception of crowds,'' \emph{IEEE
  Computer Graphics \& Applications}, vol.~31, no.~3, pp. 22--31, 2011.

\bibitem{Hu2014The}
B.~Hu, Y.~Ying, Z.~Bo, Y.~Ping, H.~Zhu, and X.~Chen, ``The sound quality
  evaluation system for recording device,'' in \emph{International Congress on
  Image and Signal Processing}, 2014.

\bibitem{50}
A.~Gupta, J.~Johnson, F.~F. Li, S.~Savarese, and A.~Alahi, ``Social
  \protect{GAN}: Socially acceptable trajectories with generative adversarial
  networks,'' in \emph{IEEE Conference on Computer Vision and Pattern
  Recognition}, 2018.

\bibitem{2}
A.~Alahi, K.~Goel, V.~Ramanathan, A.~Robicquet, and S.~Savarese, ``Social
  \protect{LSTM}: Human trajectory prediction in crowded spaces,'' in
  \emph{IEEE Conference on Computer Vision \& Pattern Recognition}, 2016.

\bibitem{Zhou2011Random}
B.~Zhou, X.~Wang, and X.~Tang, ``Random field topic model for semantic region
  analysis in crowded scenes from tracklets,'' in \emph{IEEE Conference on
  Computer Vision \& Pattern Recognition}, 2011.

\bibitem{6977426}
A.~{Bera} and D.~{Manocha}, ``Realtime multilevel crowd tracking using
  reciprocal velocity obstacles,'' in \emph{International Conference on Pattern
  Recognition}, Aug 2014, pp. 4164--4169.

\end{thebibliography}

%
\clearpage

\begin{IEEEbiography}[{\includegraphics[width=1in,height=1.25in,clip,keepaspectratio]{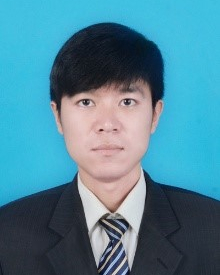}}]{Chaochao Li}
received B.S. degree in computer
science and technology and the master's degree in
computer application technology from the School
of Information Engineering, Zhengzhou University,
Zhengzhou, China, where he is currently pursuing
Ph.D. degree.

His current research interests include computer
graphics and computer vision.
\end{IEEEbiography}

\begin{IEEEbiography}[{\includegraphics[width=1in,height=1.25in,clip,keepaspectratio]{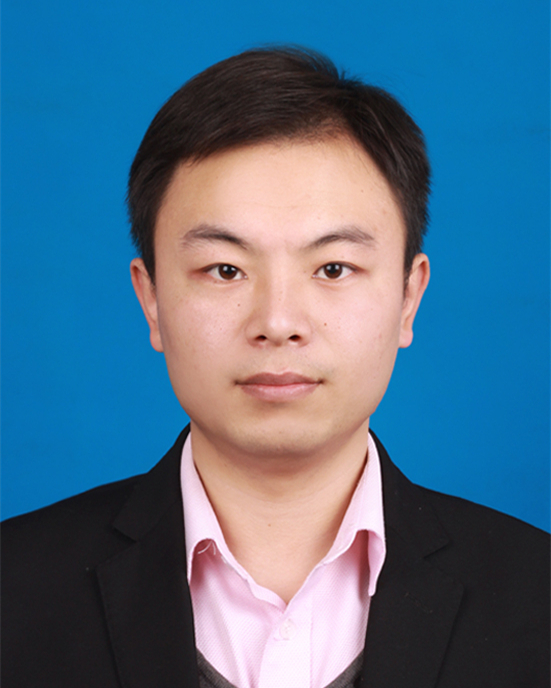}}]{Pei Lv}
received his Ph.D. degree from the State
Key Laboratory of CAD\&CG, Zhejiang University,
Hangzhou, China, in 2013.

He is currently an Associate Professor with
the School of Information Engineering, Zhengzhou
University, Zhengzhou, China. His current research
interests include video analysis and crowd simulation. He has authored over 20 journal and conference
papers in the above areas, including the IEEE
TRANSACTIONS ON IMAGE PROCESSING, IEEE
TRANSACTIONS ON CIRCUITS AND SYSTEMS FOR
VIDEO TECHNOLOGY, and ACM Multimedia.
\end{IEEEbiography}

\begin{IEEEbiography}[{\includegraphics[width=1in,height=1.25in,clip,keepaspectratio]{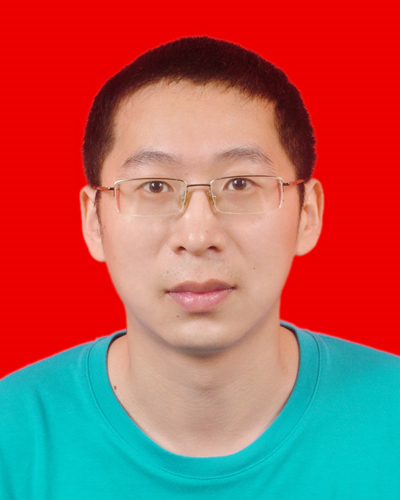}}]{Mingliang Xu}
received his Ph.D. degree in computer science and technology from the State Key Laboratory of CAD\&CG, Zhejiang University, Hangzhou, China.

He is a Full Professor with the School of Information Engineering, Zhengzhou University, Zhengzhou, China, where he is currently the Director of the Center for Interdisciplinary Information Science Research. He was with the Department of Information Science, National Natural Science Foundation of China, Beijing,
China, from 2015 to 2016. His current research interests include computer
graphics, multimedia, and artificial intelligence. He has authored over 60
journal and conference papers in the above areas, including the ACM
Transactions on Graphics, ACM Transactions on Intelligent Systems and
Technology, IEEE TRANSACTIONS ON PATTERN ANALYSIS AND MACHINE
INTELLIGENCE, IEEE TRANSACTIONS ON IMAGE PROCESSING, IEEE
TRANSACTIONS ON CYBERNETICS, IEEE TRANSACTIONS ON CIRCUITS
AND SYSTEMS FOR VIDEO TECHNOLOGY, ACM SIGGRAPH (Asia), ACM
MM, and ICCV.

Dr. Xu is the Vice General Secretary of ACM SIGAI China.
\end{IEEEbiography}

\begin{IEEEbiography}[{\includegraphics[width=1in,height=1.25in,clip,keepaspectratio]{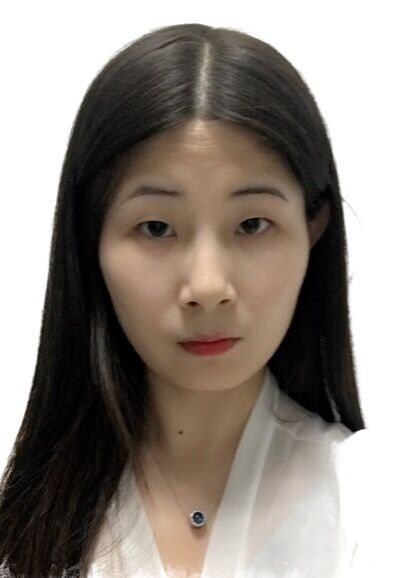}}]{Xinyu Wang}
is a graduate student in the School of Information Engineering of Zhengzhou University, China. Her research interests include pedestrian trajectory prediction and crowd simluation.
\end{IEEEbiography}

\begin{IEEEbiography}[{\includegraphics[width=1in,height=1.25in,clip,keepaspectratio]{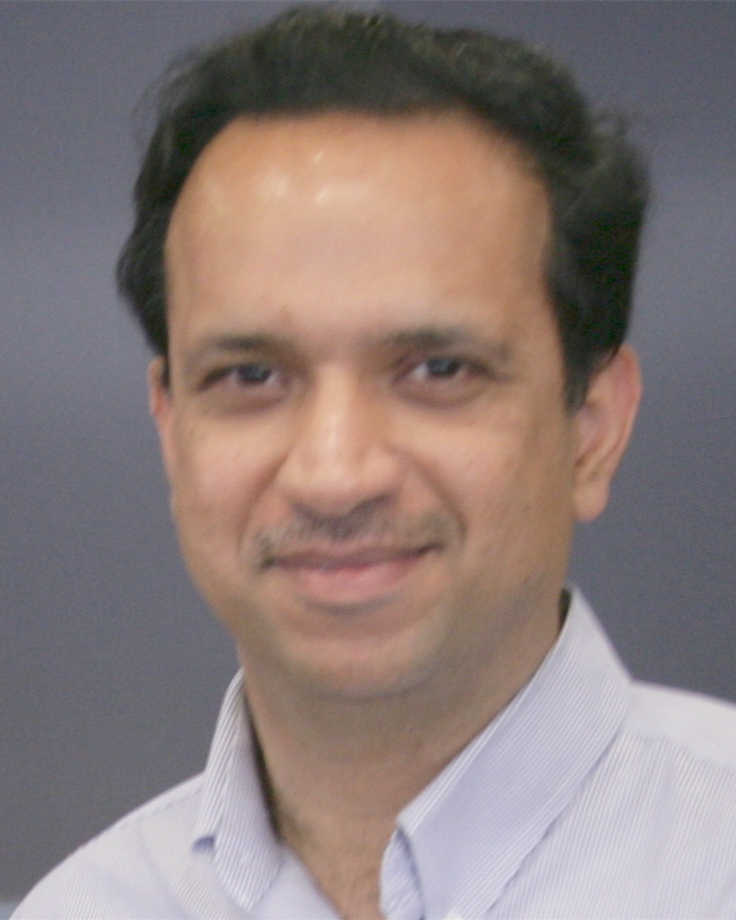}}]{Dinesh Manocha}
is the Paul Chrisman Iribe Chair in Computer Science \& Electrical and Computer Engineering at the University of Maryland College Park. He is also the Phi Delta Theta/Matthew Mason Distinguished Professor Emeritus of Computer Science at the University of North Carolina - Chapel Hill. He has won many awards, including Alfred P. Sloan Research Fellow, the NSF Career Award, the ONR Young Investigator Award, and the Hettleman Prize for scholarly achievement. His research interests include multi-agent simulation, virtual environments, physically-based modeling, and robotics. His group has developed a number of packages for multi-agent simulation, crowd simulation, and physics-based simulation that have been used by hundreds of thousands of users and licensed to more than 60 commercial vendors. He has published more than 500 papers and supervised more than 35 PhD dissertations. He is an inventor of 9 patents, several of which have been licensed to industry. His work has been covered by the New York Times, NPR, Boston Globe, Washington Post, ZDNet, as well as DARPA Legacy Press Release. He is a Fellow of AAAI, AAAS, ACM, and IEEE and also received the Distinguished Alumni Award from IIT Delhi. He was a co-founder of Impulsonic, a developer of physics-based audio simulation technologies, which was acquired by Valve Inc. See \url{http://www.cs.umd.edu/~dm}
\end{IEEEbiography}

%
\begin{IEEEbiography}[{\includegraphics[width=1in,height=1.25in,clip]{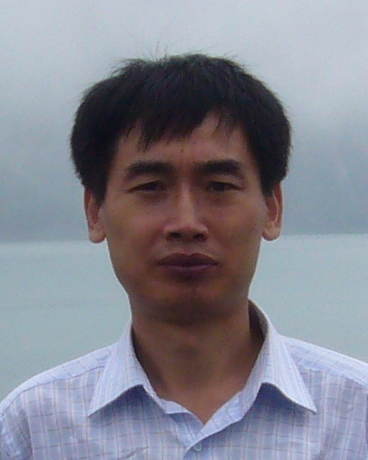}}]{Bing Zhou}
received B.S. and M.S. degrees from
Xi'an Jiaotong University, Xi'an, China, in 1986
and 1989, respectively, and a Ph.D. degree from
Beihang University, Beijing, China, in 2003, all in
computer science.

He is currently a Professor with the School
of Information Engineering, Zhengzhou University,
Zhengzhou, China. His current research interests
include video processing and understanding, surveillance, computer vision, and multimedia applications.
\end{IEEEbiography}
\begin{IEEEbiography}[{\includegraphics[width=1in,height=1.25in,clip]{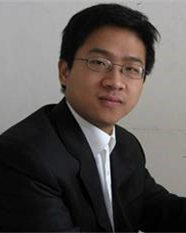}}]{Meng Wang}
received B.S. and Ph.D. degrees in signal and information processing
from the University of Science and Technology
of China, Hefei, China, in 2003 and 2008, respectively.

He is a Professor with the Hefei University of
Technology, Hefei. His current research interests
include multimedia content analysis, computer
vision, and pattern recognition. He has authored over
200 book chapters, journal and conference papers in
the above areas.

Dr. Wang was a recipient of the ACM SIGMM Rising Star Award 2014.
He is an Associate Editor of the IEEE TRANSACTIONS ON KNOWLEDGE
AND DATA ENGINEERING, IEEE TRANSACTIONS ON CIRCUITS AND
SYSTEMS FOR VIDEO TECHNOLOGY, and IEEE TRANSACTIONS ON
NEURAL NETWORKS AND LEARNING SYSTEMS.
\end{IEEEbiography}


%
%
%




\end{document}